\documentclass[%
 reprint,
superscriptaddress,
nofootinbib,
 amsmath,amssymb,
 aps,
]{revtex4-1}

\usepackage{amsmath,amssymb}
\usepackage{graphicx}
\usepackage{dcolumn}
\usepackage{bm}
\usepackage{hyperref}

\begin{document}

\title{Fermion-boson stars with a quartic self-interaction in the boson sector}

\author{Susana Valdez-Alvarado} 
 \email{svaldez@fisica.ugto.mx}
\affiliation{%
Facultad de Ciencias de la Universidad Aut\'onoma del Estado de M\'exico  (UAEM\'ex.), Instituto Literario No. 100, C.P. 50000, Toluca, Estado de M\'exico, M\'exico}

\author{Ricardo Becerril}%
 \email{becerril@ifm.umich.mx}
 \affiliation{Instituto de F\'isica y Matem\'aticas, Universidad Michoacana de San Nicol\'as de Hidalgo. Edif. C-3, 58040 Morelia, Michoac\'an, M\'exico}

\author{L. Arturo Ure\~{n}a-L\'{o}pez} 
 \email{lurena@ugto.mx}
\affiliation{%
Departamento de F\'isica, DCI, Campus Le\'on, Universidad de
Guanajuato, 37150, Le\'on, Guanajuato, M\'exico.}

\begin{abstract}
  Fermion-boson stars are solutions of the gravitationally coupled Einstein-Klein-Gordon-Hydrodynamic equations system. By means of  methods developed in previous works, we perform a stability analysis of fermion-boson stars that include a quartic self-interaction in their bosonic part. Additionally, we describe the complete structure of the stability and instability regions of the space of free parameters, which we argue is qualitatively the same for any value of the quartic self-interaction. The relationship between the total mass of mixed stars and their general stability is also discussed in terms of the structure identified within the stability region.
\end{abstract}

\maketitle

\section{Introduction \label{sec:introduction}}

Self-gravitating systems are one of the most interesting areas in gravitation, astrophysics and cosmology, mainly because of the characteristic features of this kind of systems that emerge directly from the nature of their main matter constituents. Fermionic self-gravitating objects are the most studied ones, as they help us to understand the properties of stars, given that hydrogen is the most abundant chemical element in the Universe~\cite{Aghanim:2018eyx,Fields:2019pfx}.

However, one may also consider the formation of bosonic self-gravitating systems, being boson stars some of the simplest ones. From the first studies in~\cite{Ruffini:1969qy}, their intrinsic characteristics have also been widely studied~\cite{Jetzer:1991jr,Schunck:2003kk,Liebling:2012fv}, and their appealing has been recently renewed because of the possibility of explaining the nature of the dark matter in the Universe by means of ultra-light bosons, see for instance the reviews in~\cite{Matos:2008ag,Magana:2012xe,Marsh:2015xka,Hui:2016ltb,Urena-Lopez:2019xri}. 

Given, on one hand, the proved existence of self-gravitating fermions, and on the other hand, the possible presence of self-gravitating bosons, it was just natural to ask for the properties of self-gravitating objects with a mixing of fermions and bosons, which have been known ever since as fermion-boson stars, see the seminal papers in~\cite{Henriques:1989ar,Henriques:1989ez,Lopes:1992aw,Henriques:2003yr}. As with any other self-gravitating system, the central question is whether fermion-boson stars are stable. This is not a trivial question at all, as fermion-boson stars are described by two parameters, and then the stability criterions employed for purely fermion o purely boson stars, which are each a one-parameter family of solutions, are no longer valid. 

Nonetheless, the authors in~\cite{Henriques:1989ez} were able to establish general guidelines to study the stability of mixed stars, and finding the corresponding region of stability on the two-parameter plane of equilibrium configurations. This was taken as a starting point in~\cite{ValdezAlvarado:2012xc}, where the dynamical evolution of fermion-boson stars was studied. The case considered was a fermion star (neutron star modeled as a perfect fluid), mixed with a boson component modeled with a scalar field, the latter being endowed with only a quadratic potential. 

The analysis performed in~\cite{ValdezAlvarado:2012xc} was based on the behaviour of the bosonic and fermionic particle numbers in mixed configurations with a fixed total mass. Following the guidelines in~\cite{Henriques:1989ez}, one looks for the maximum/minimum of the curves associated to the particle numbers, and thereby one can distinguish stable configurations from the unstable ones. The foregoing stability criterion was confirmed by numerically evolving some equilibrium configurations.

In this work, the stability analysis is presented for fermion-boson
stars with a quartic self-intertaction in the bosonic part, using the methodology developed in~\cite{ValdezAlvarado:2012xc}. Specifically, we study the influence of the self-interaction term on the total mass, the size, and the number of bosonic and fermionic particles of mixed stars. We will also extend the study of~\cite{Vallisneri:1999nq} and establish the structure of the stability and instability regions.

This paper is organized as follows. In Sec.~\ref{sec:formalism}, we present the equations of motion and their boundary conditions that allow the construction of equilibrium configuration for mixed stars. Then, Sec.~\ref{sec:stability-analysis} is devoted to the study of the stability of the mixed stars, and a general discussion on the structure of the two-parameters plane of equilibrium configurations. Finally, Sec.~\ref{sec:final-rem} presents the conclusions and final remarks.

\section{Mathematical Formalism \label{sec:formalism}} 

Boson-Fermion stars can be modeled by a complex scalar field $\phi$ endowed 
with a scalar potential $V(|\phi|)$, which will represent the bosonic
part, and a perfect fluid, which is described by the following primitive physical variables: the rest-mass density $\rho$, the pressure $P$, the internal energy $\epsilon$ and its 4-velocity $u^{\mu}$. 

The equations of motion are
the coupled Einstein-Klein-Gordon-Hydrodynamic equations,
given by
\begin{subequations}
\label{eq:ekg} 
\begin{eqnarray}
 G_{\mu\nu} &=& 8\pi \left(T^{(\phi)}_{\mu\nu} + T^{(f)}_{\mu\nu} \right)
  , \quad \Box^2\phi -\phi\frac{dV(\phi)}{d(|\phi|^2)}=0 \, , \label{eq:ekg-a} \\
& & \,\,\, \nabla_\mu T^{(f) \mu\nu}=0 \,\,\, \,\,\,, \quad \nabla_{\mu}(\rho
u^{\mu})=0 \, . \label{eq:ekg-b}
\end{eqnarray}
\end{subequations}
where we use geometric units in which $G=c=1$. Here, $T^{(\phi)}_{\mu \nu}$ and $T^{(f)}_{\mu \nu}$ are the stress-energy tensors of the bosonic and fermionic components, respectively, which are explicitly given by

\begin{eqnarray}
T^{\phi}_{\mu\nu} &=& \frac{1}{2} \left( \partial_{\mu}\phi^*\partial_{\nu}\phi+ 
\partial_{\mu}\phi\partial_{\nu}\phi^* \right) -\frac{g_{\mu\nu}}{2}
\left( \partial^{\alpha}\phi^*\partial_{\alpha}\phi+2V \right) \, ,\nonumber  \\
T^{f}_{\mu\nu}  &=& \left[\rho(1+\epsilon)+P \right]u_{\mu}u_{\nu}+Pg_{\mu\nu} \, .
\label{eq:hd}
\end{eqnarray}

The scalar field potential is written as
\begin{equation}
V(\Phi) = \frac{m^2}{2} |\phi|^2 + \frac{\lambda}{4} |\phi|^4 \, ,
\label{eq:potential}
\end{equation}
which represents boson particles with mass $m$ and a self-interaction parameter $\lambda$. 

\subsection{Evolution equations
\label{sec:evolution-equations}}
The equations of motion of mixed stars with self-interacting bosons are obtained by considering the time-dependent spherically symmetric metric, 
\begin{equation}
    ds^2 = -\alpha^2(t,r)dt^2 + g_{rr}(t,r)dr^2 + r^2g_{\theta\theta} d\Omega^2 \, ,
\end{equation}
and the potential (\ref{eq:potential}) into the equations (\ref{eq:ekg-a}) and (\ref{eq:ekg-b}). The evolution equations of the scalar field  and perfect fluid, are 
\begin{widetext}
\begin{subequations}
\label{eq:ekghd}
\begin{eqnarray}
\partial_t\phi_t &=& \partial_r(\alpha\sqrt{g^{rr}}\phi_r)+\alpha\sqrt{g^{rr}}\Big[2\Big(D_{r\theta}{}^{\theta}+
\frac{1}{r}\Big)\phi_r+2\sqrt{g_{rr}}K_{\theta}{}^{\theta}\phi_t-g_{rr}\phi(m^2+\lambda|\phi|^2)\Big],
\\
\partial_t(\sqrt{\gamma}D) &=& -\partial_r(\sqrt{\gamma}\alpha
v^{r}D) - \frac{2}{r}\sqrt{\gamma}\alpha v^rD,
\\
\partial_t(\sqrt{\gamma}U) &=& -\partial_r(\sqrt{\gamma}\alpha
\tilde{S}^r)+\sqrt{\gamma}\alpha\Big[\tilde{S}_r{}^rK_r{}^r+2\tilde{S}_{\theta}{}^{\theta}K_{\theta}{}^{\theta}-
\tilde{S}^r\Big(\frac{2}{r}+A_r\Big)\Big],
\\
\partial_t(\sqrt{\gamma}\tilde{S}_r) &=& -\partial_r(\sqrt{\gamma}\alpha
\tilde{S}_r{}^r)+\sqrt{\gamma}\alpha\Big[\tilde{S}_r{}^r\Big(D_{rr}{}^r-\frac{2}{r}\Big)+
2\tilde{S}_{\theta}{}^{\theta}\Big(\frac{1}{r}+D_{r\theta}^{\theta}\Big)-UA_r\Big], 
\end{eqnarray}
\end{subequations}
\end{widetext}
where $\sqrt{\gamma}=\sqrt{g_{rr}}g_{\theta\theta}$ and $A_r$, $D_{rr}^{r}$, $D_{r\theta}^{\theta}$, $K_r^r$ and $K_{\theta}^{\theta}$ are variables defined in the $Z3$ formulation of the Einstein equations that we use to describe the evolution of the space-time~\cite{Alic:2007ev}. Appendix~\ref{app:Einsten-equations} contains the description of all these variables and the complete set of equations of motion obtained with this formulation. 

To reduced the Klein-Gordon equation to first order in space and time, we were introduced the following auxiliary variables 
\begin{eqnarray}
\phi_r &=& \partial_r\phi\, , \,\,\,
\phi_t=\frac{\sqrt{g_{rr}}}{\alpha}\partial_t\phi\,.
\end{eqnarray}
The mass density, $D$, the momentum density $\tilde{S}$ and the energy density, $U$, are used to described the evolution of the perfect fluid, which are known as conserved variables. These conserved variables are related with the primitive variables: the rest mass density, $\rho_0$, the pressure $p$ and the velocity $v_r$. In Appendix~\ref{app:conserved-primitive} we will define the conserved variables and the spatial projections of the stress-energy tensor, $\tilde{S}_r
^r$ and $\tilde{S}_{\theta}^{\theta}$, in function of the primitive variables. As well, we indicate the process to obtain the primitive variables in terms of the conserved ones. 

\subsection{Equilibrium Configurations \label{sec:equilibrium-conf}}

We shall be interested in equilibrium configurations, for which we assume a static and spherically symmetric metric in the form
\begin{subequations}
\begin{equation}
ds^2 = -\alpha^2(r)dt^2 + a^2(r)dr^2 + r^2 d\Omega^2 \, . \label{eq:metric}
\end{equation}
For the complex scalar field, we assume the standard harmonic form $\phi(t,r)=\phi(r)e^{-i\omega t}$, where 
$\omega$ is an intrinsic frequency, whereas for the perfect fluid in hydrostatic equilibrium, we take the following four-velocity $u^{\mu}=(-1/\alpha,0,0,0)$. For numerical purposes, we consider the following set of new dimensionless variables: 
\begin{eqnarray}
x= mr\, , \quad  \Omega=\frac{\omega}{m}\, , \quad \sqrt{4\pi} \phi \to \phi \, , \\
\Lambda = \frac{m^2_{Pl} \lambda}{4\pi m^2}\, , \quad \frac{4\pi}{m^2} \rho \to \rho \, ,  \quad \frac{4\pi}{m^2}P \to P \, ,
\end{eqnarray}
\end{subequations}
Thus, the equations of motion~\eqref{eq:ekg} for equilibrium configurations explicitly read
\begin{subequations}
\begin{widetext}
\begin{eqnarray}
a^\prime &=& \frac{a}{2}\left(\frac{1-a^2}{x} +
  a^2x\left[\left(\frac{\Omega^2}{\alpha^2} + 1 + \frac{\Lambda}{2}\phi^2
  \right)\phi^2 + \frac{\Phi^2}{a^2} + 2\rho(1+\epsilon)\right] \right),\\
\alpha^\prime &=& \frac{\alpha}{2}\left(\frac{a^2-1}{x} +
  a^2x\left[\left(\frac{\Omega^2}{\alpha^2} - 1
  -\frac{\Lambda}{2}\phi^2\right)\phi^2 + \frac{\Phi^2}{a^2} +
  2P\right] \right),\\
\phi^\prime &=& \Phi, \\
\Phi^\prime &=& \left(1 - \frac{\Omega^2}{\alpha^2} + \Lambda \phi^2 \right) a^2 \phi -
  \left(\frac{2}{x} + \frac{\alpha^\prime}{\alpha} -
                     \frac{a^\prime}{a}\right) \Phi, \\
P^\prime &=& -\frac{\alpha^\prime}{\alpha} [\rho(1+\epsilon) + P] \, ,
\end{eqnarray}
\label{eq:ekghd}
\end{widetext}
\end{subequations}
where a prime denotes derivative with respect to $x$. In order to obtain a closed  system, we must introduce an equation of state for the perfect fluid. Particularly, we adopt a polytropic equation of state $P=K \rho^{\Gamma}$, with polytropic constant $K=100$ and adiabatic index $\Gamma=2$, which corresponds to masses and compactness in the range of neutron stars~\cite{palenzuelabook}\footnote{\bf Modern studies of neutron stars consider more involved models, see for instance
~\cite{Abbott_2018}. Our choice here is for purposes of simplicity and for a better comparison with the results in~\cite{ValdezAlvarado:2012xc}. The study of more general parametrizations of the equation of state is left for future work and will be presented elsewhere.}.

The system of equations~\eqref{eq:ekghd} represents an eigenvalue problem for the frequency of the bosonic part
$\Omega$, as a function of the central values of the fluid density $\rho_0$ and the scalar field $\phi_0$. We solve this system
by using the shooting method \cite{nr} and under boundary conditions corresponding to regularity at the origin and asymptotic flatness at infinity. The latter are:
\begin{subequations}
\begin{eqnarray}
a(0) &=& 1\, , \quad \lim \limits_{x \rightarrow \infty} a(x)=1\, , \\
\alpha(0) &=&1 \, , \quad \lim \limits_{r \rightarrow \infty} \alpha(x) =
  \lim \limits_{x \rightarrow \infty} \frac{1}{a(x)} \, , \\
\phi(0) &=& \phi_0 \,  , \quad \lim \limits_{x \rightarrow \infty} 
           \phi(x) = 0\, , \quad \Phi(0)=0\, , \\
\rho(0) &=& \rho_0 \, , \quad P(0) = K \rho^{\Gamma}_0 \, , \quad \lim \limits_{x \rightarrow \infty} P(x)=0. 
\end{eqnarray}
\end{subequations}

Additionally, we use the Schwarzschild mass to obtain the mass of equilibrium configurations,
\begin{equation}
M _T =  \lim \limits_{x \rightarrow \infty} \frac{x}{2} \left(1 -\frac{1}{\alpha^2}\right). 
\end{equation}
Also, due to the symmetry $U(1)$ in the Lagrangian of the scalar field and the Noether theorem, the  scalar field charge is conserved, which can
be associated with the number of bosons $N_B$, whereas the number of fermions $N_F$ is defined by the conservation 
of the baryonic number. $N_B$ and $N_F$ are then calculated by means of the following expressions,
\begin{equation}
\frac{\partial N_B}{\partial r} = \frac{4\pi
  a\omega\phi^2r^2}{\alpha}\, , ~~~ ~~~ \frac{\partial N_F}{\partial r} =
4\pi a\rho r^2.
\end{equation}
Finally, we define the radius of a star $R_T$ as the value of $r$ containing $95\%$ of the total
mass. Correspondingly, the radius of the bosonic (fermionic) component $R_B$ ($R_F$), will be the value of $r$ containing $95\%$ of the corresponding
particles.

\subsection{Numerical solutions \label{sec:numerical-sol}}

\begin{figure}[ht]
  \begin{center}
  \includegraphics[width=0.43\textwidth]{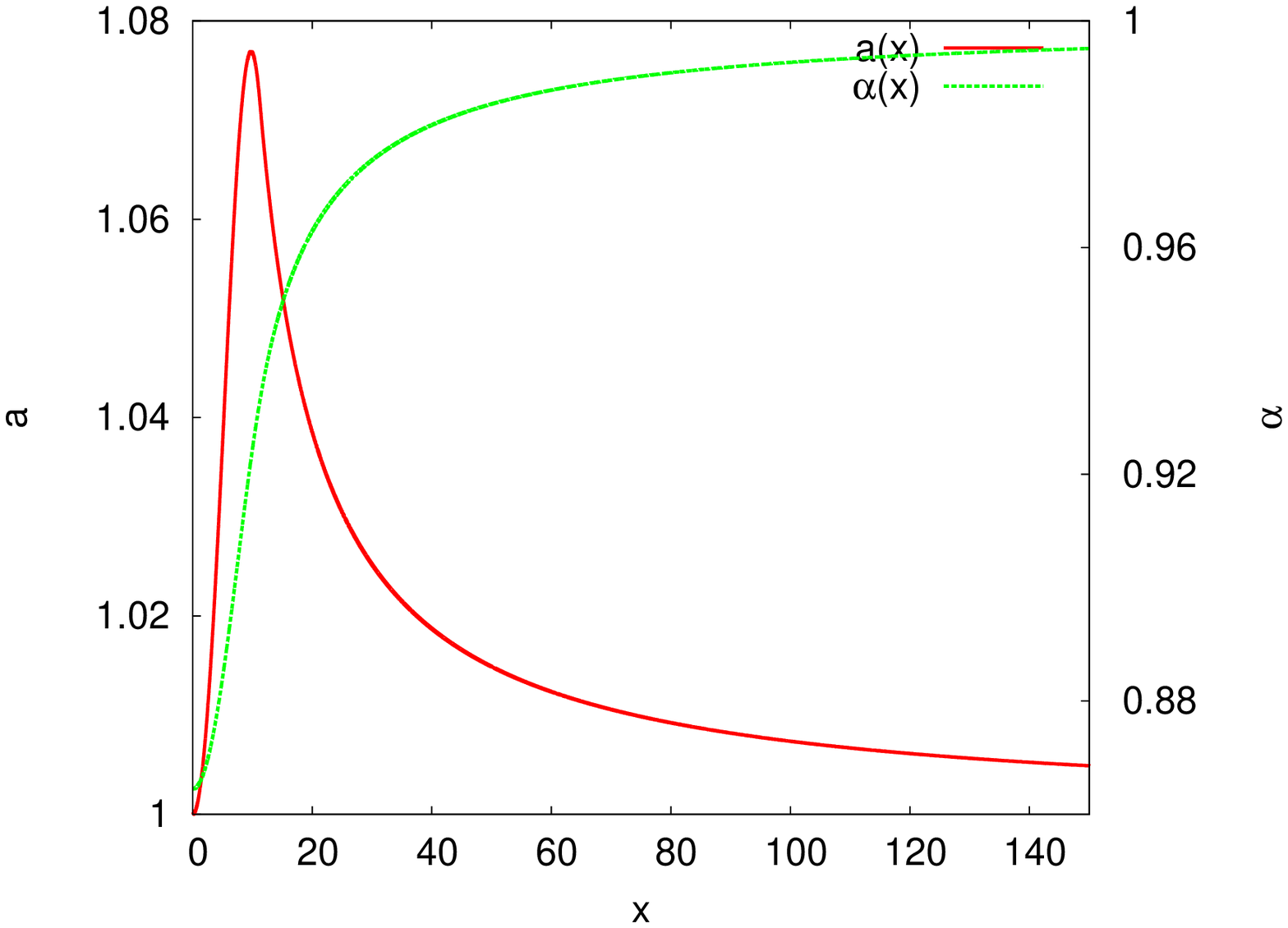}
  \includegraphics[width=0.43\textwidth]{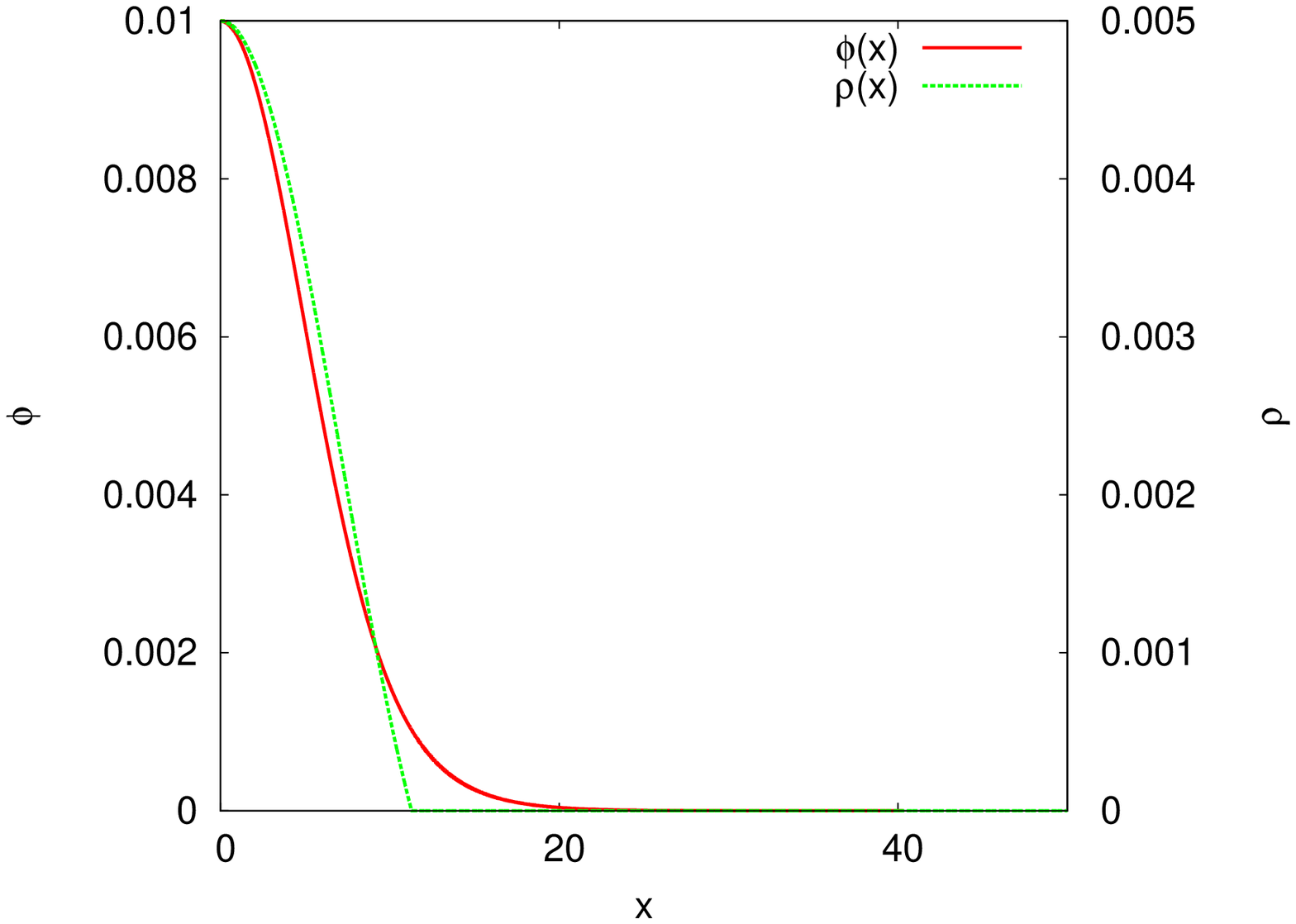}
      \caption{(Top) Numerical solutions of the metric functions $a(x)$ and $\alpha(x)$, see Eq.~\eqref{eq:metric}, which are non-singular at the origin and asymptotically flat at infinity. (Bottom) Numerical solutions of the scalar field $\phi(x)$ and the fluid density $\rho(x)$, which are regular at the origin and vanish at infinity. The profiles correspond to a configuration built from $\phi_0=0.01$, $\rho_0= 0.005$ and $\Lambda=10$.}
      \label{conditions}
  \end{center}
\end{figure}

In this section, we construct the equilibrium configurations of the mixed
boson-fermion stars by solving numerically the system of equations 
\eqref{eq:ekghd}. In Fig.~\ref{conditions}, we can see a typical example, which satisfies our conditions of regularity at the origin and asymptotical flatness at infinity. This particular configuration was constructed with a central scalar field  $\phi(0)=\phi_0=0.01$, a central fluid density $\rho_0= 0.005$ and $\Lambda=10$.

\begin{figure}[h!]
  \begin{center}
  \includegraphics[width=0.43\textwidth]{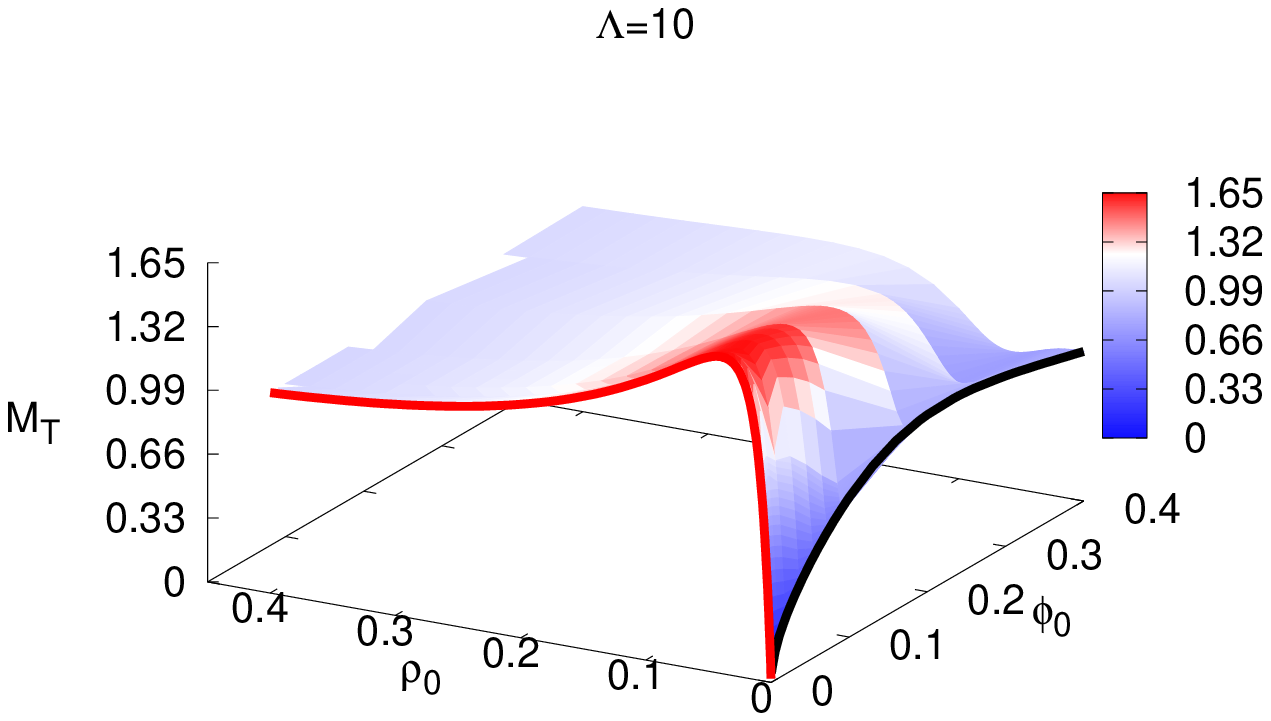}
\caption{The total mass of mixed stars as a function of the central values, $M_T=M_T(\rho_c,\phi_c)$, here for the self-interaction parameter $\Lambda = 10$. On the plane $\rho_0 = 0$, one can see the typical
  curve (in black) $M_B=M_B(\phi_0)$ for a purely-boson star, whereas on the plane $\phi_0=0$ one can see the typical curve (in red) $M_F=M_F(\rho_0)$ for a purely-fermion star.}
\label{totalmass}
  \end{center}
\end{figure}

Fig.~\ref{totalmass} shows the total mass of fermion-boson
stars configurations as
a function of $\rho_0$ and $\phi_0$, that is
$M_T=M_T(\rho_0,\phi_0;\Lambda)$, with a self-interaction parameter
$\Lambda=10$. In the plane $\rho_0 = 0$, one may see the typical
curve $M_B=M_B(\phi_0)$ for a purely boson star, whereas in the
plane $\phi_0=0$, one may see the typical
curve $M_F=M_F(\rho_0)$ of a purely fermion star.

It is possible to verify that, if we set
$\phi_0=0$, then we get back to the mass curve of neutron stars
with a critical mass $M_{Fc}=1.64$, that corresponds to $K=100$ and $\Gamma=2$.
This mass separates the stable configurations
($ M_T< M_{Fc}$) from the unstable ones ($M_T>M_{Fc}$) in the case of purely-fermion stars. On the other hand,
if we set $\rho_0=0$, we recover the mass curve of boson stars,
which have a critical mass $M_{Bc}=0.92$ corresponding to the value $\Lambda=10$ \cite{Colpi:1986ye}.

In Table~\ref{t1} we show the resultant numerical values of different equilibrium configurations obtained from Eqs.~\eqref{eq:ekghd}, taking into account the self-interaction in the bosonic part: $\Lambda=0,10,30$. We also indicate the stability of each numerical case, the details of which we explain in Sec.~\ref{sec:stability-analysis} below.

\begin{table*}
\begin{tabular}{|rr|rrrr|rrrr|rrrrrrr|}
\hline
\multicolumn{2}{|c|}{$\Lambda=0$} & \multicolumn{4}{|c|}{Fermion Star $(\rho_0 \neq 0,\phi_0=0)$} & \multicolumn{4}{|c|}{Boson Star $(\rho_0=0,\phi_0 \neq 0)$} & \multicolumn{7}{|c|}{Mixed Star $(\rho_0 \neq 0,\phi_o \neq 0)$}\\
\hline
\hline
$\rho_0$ & $\phi_0$ & $M_F$ & $R_F$ & $N_F$ & & $M_B$ & $R_B$ & $N_B$ & & $M_T$ & $R_{T}$ & $R_F$ & $R_B$ & $N_F$ & $N_B$\\

\hline
 & $0.01$ & & & & & $0.202$ & $37.662$ & $0.202$ & stable & $1.636$ & $7.20$ & $6.757$ & $4.587$ & $1.800$ & $7.07 (10^{-4})$ & stable\\
$0.04$ & $0.27$ & $1.637$ & $6.750$ & $1.798$ & stable & $0.633$ & $5.929$ & $0.653$ & stable & $1.035$ & $6.33$ & $5.931$ & $4.223$ & $0.784$ & $0.329$ & stable \\
& $0.60$ &  & & & & $0.525$ & $3.335$ & $0.516$ & unstable & $0.513$ & $4.68$ & $1.455$ & $3.301$ & $0.016$ & $0.489$ & unstable\\

\hline
 & $0.01$ & & & & & $0.202$ & $37.662$ & $0.202$ & stable & $1.594$ & $6.36$ & $5.955$ & $3.818$ & $1.740$ & $4.33 (10^{-4})$ & unstable\\
$0.06$ & $0.27$ & $1.594$ & $5.947$ & $1.739$ & unstable & $0.633$ & $5.929$ & $0.653$ & stable & $1.255$ & $6.22$ & $5.814$ & $3.611$ & $1.130$ & $0.234$ & unstable \\
 & $0.60$ & & & & & $0.525$ & $3.335$ & $0.516$ & unstable & $0.508$ & $4.46$ & $2.085$ & $3.170$ & $0.053$ & $0.448$ & unstable\\

\hline
\multicolumn{2}{|c|}{$\Lambda=10$} & \multicolumn{4}{|c|}{} & \multicolumn{4}{|c|}{} & \multicolumn{7}{|c|}{}\\
\hline
$\rho_0$ & $\phi_0$ & $M_F$ & $R_F$ & $N_F$ & & $M_B$ & $R_B$ & $N_B$ & & $M_T$ & $R_{T}$ & $R_F$ & $R_B$ & $N_F$ & $N_B$\\

\hline
 & $0.01$ & & & & & $0.210$ & $37.846$ & $0.211$ & stable & $1.636$ & $7.20$ & $6.757$ & $4.587$ & $1.800$ & $7.08 (10^{-4})$ & stable\\
$0.04$ & $0.22$ & $1.637$ & $6.750$ & $1.798$ & stable & $0.920$ & $6.796$ & $0.964$ & stable & $1.084$ & $6.13$ & $5.716$ & $4.665$ & $0.799$ & $0.370$ & stable\\
 & $0.35$ &  & & & & $0.830$ & $4.835$ & $0.852$ & unstable & $0.758$ & $5.94$ & $2.825$ & $4.537$ & $0.116$ & $0.665$ & unstable\\
\hline
 & $0.01$ &  &  &  & & $0.210$ & $37.846$ & $0.211$ & stable & $1.594$ & $6.36$ & $5.955$ & $3.818$ & $1.740$ & $4.33(10^{-4})$ & unstable\\
$0.06$ & $0.22$ & $1.594$ & $5.951$ & $1.739$ & unstable & $0.920$ & $6.796$ & $0.964$ & stable & $1.288$ & $6.11$ & $5.708$ & $3.881$ & $1.170$ & $0.232$ & unstable\\
 & $0.35$ &  &  &  & & $0.830$ & $4.835$ & $0.852$ & unstable & $0.796$ & $4.86$ & $3.902$ & $3.925$ & $0.329$ & $0.500$ & unstable\\
\hline
\multicolumn{2}{|c|}{$\Lambda=30$} & \multicolumn{4}{|c|}{} & \multicolumn{4}{|c|}{} & \multicolumn{7}{|c|}{}\\
\hline
$\rho_0$ & $\phi_0$ & $M_F$ & $R_F$ & $N_F$ & & $M_B$ & $R_B$ & $N_B$ & & $M_T$ & $R_{T}$ & $R_F$ & $R_B$ & $N_F$ & $N_B$\\

\hline
 & $0.01$ & & & & & $0.228$ & $38.178$ & $0.229$ & stable & $1.636$ & $7.20$ & $6.757$ & $4.591$ & $1.800$ & $7.10(10^{-4})$ & stable\\
$0.04$ & $0.16$ & $1.637$ & $6.750$ & $1.798$ & stable & $1.336$ & $8.652$ & $1.413$ & stable & $1.255$ & $6.42$ & $6.000$ & $5.144$ & $1.050$ & $0.315$ & stable\\
 & $0.25$ &  &  &  &  & $1.177$ & $6.244$ & $1.215$ & unstable & $0.989$ & $7.11$ & $3.463$ & $5.671$ & $0.228$ & $0.801$ & unstable\\
\hline
 & $0.01$ & & & & & $0.228$ & $38.178$ & $0.229$ & stable & $1.594$ & $6.36$ & $5.954$ & $3.819$ & $1.740$ & $4.34(10^{-4})$ & unstable\\
$0.06$ & $0.16$ & $1.594$ & $5.947$ & $1.739$ & unstable & $1.336$ & $8.652$ & $1.413$ & stable & $1.397$ & $6.17$ & $5.765$ & $4.135$ & $1.350$ & $0.170$ & unstable\\
 & $0.25$ &  & &  & & $1.177$ & $6.244$ & $1.215$ & unstable & $0.990$ & $5.32$ & $4.355$ & $4.560$ & $0.533$ & $0.510$ & unstable\\
\hline
\hline
\end{tabular}
\caption{Properties of fermion, boson and mixed stars. The columns report, from left to right: the central density of the perfect fluid $\rho_0$, the central value of the scalar field $\phi_0$, the mass $M_F$, radius $R_F$ and particles number $N_F$ of the purely-fermion stars; the mass $M_B$, radius $R_B$ and particles number $N_B$ of the purely-boson stars; the total mass $M_T$, total radius $R_{T}$, the radii $R_F$, $R_B$ and particle numbers $N_F$, $N_B$ of the fermionic and bosonic components, respectively, of the mixed stars. See the text for details.}
\label{t1}
\end{table*}

In the first row of the first column of this table (for $\Lambda=0$)
we have one value of $\rho_0=0.04$ employed to
construct a purely stable fermion star (whose mass $M_F$, radius $R_F$ and
number of particles $N_F$ are displayed in the second column),
and three values of $\phi_0=0.01, 0.271, 0.6$
employed to construct three purely boson stars, two stable and one unstable
(whose masses $M_B$, radius $R_B$ and
number of particles $N_B$, are displayed in the third column). 

In the fourth
column, features of three mixed stars formed from
$(\rho_0,\phi_0)=(0.04,0.01), (0.04,0.271)$ and $(0.04,0.6)$ are displayed.
As long as both individual stars (fermionic and bosonic type) are stable, the
corresponding mixed star is stable,  but it is unstable if either individual star is not stable (as it is
shown in the second row where $\rho_0=0.06$ and $\phi_0=0.01, 0.27, 0.6$ were used). We observe that $M_T \leq Max\{M_F,M_B\}$ that is to say, the total mass of a mixed star is typically smaller than the maximum value
between $M_F$ and $M_B$ of individual stars.

Results for $\Lambda=10,30$ are also presented in Table \ref{t1}.
For a fixed value of $\Lambda$, when a mixed star is formed from individual stable stars (purely
fermionic/bosonic) its characteristics are closer to those of the more
massive individual star (either fermionic or bosonic). However, if a mixed
star is formed from a bosonic unstable star regardless what the value
of $M_F$ is, its characteristics are closer to those of the individual
boson star. For $\Lambda \neq 0$, $M_T \leq Max\{M_F,M_B\}$ still stands.

\section{Stability analysis \label{sec:stability-analysis}}

Because mixed stars are parameterized by the two quantities $\phi_0$ and $\rho_0$, we can not use the stability theorems for single parameter solutions to carry out the stability analysis of these stars. Then, to study the stability of mixed configuration we use the method developed in\cite{ValdezAlvarado:2012xc} that is, the stability analysis is carried out examining the behaviour of fermion and boson number, yet fixing the mass value. Thereupon, the stability curve is formed with the pair ($\rho_0$, $\phi_0$) exactly in the point where the number of particles reached the minimum and maximum values.

\subsection{General stability regions \label{sec:general-stab}}

The foregoing method is based on the behavior of $N_B$ and $N_F$, for configurations with the same value of $M_T$. Beginning with a purely fermionic star (by providing $\rho_0$ and setting $\phi_0 =0$), mixed configurations are then built up by increasing the value of $\phi_0$ from zero in such a way that $M_T$ is kept fixed. From the curve of the number of particles in terms of $\phi_0$ and $\rho_0$, it can be observed that $N_F$ decreases to a minimum while $N_B$
increases until reaching a maximum, at exactly the same values of $\phi_0$ and $\rho_0$. 

\begin{figure}[ht!]
  \begin{center}
  \includegraphics[width=0.45\textwidth]{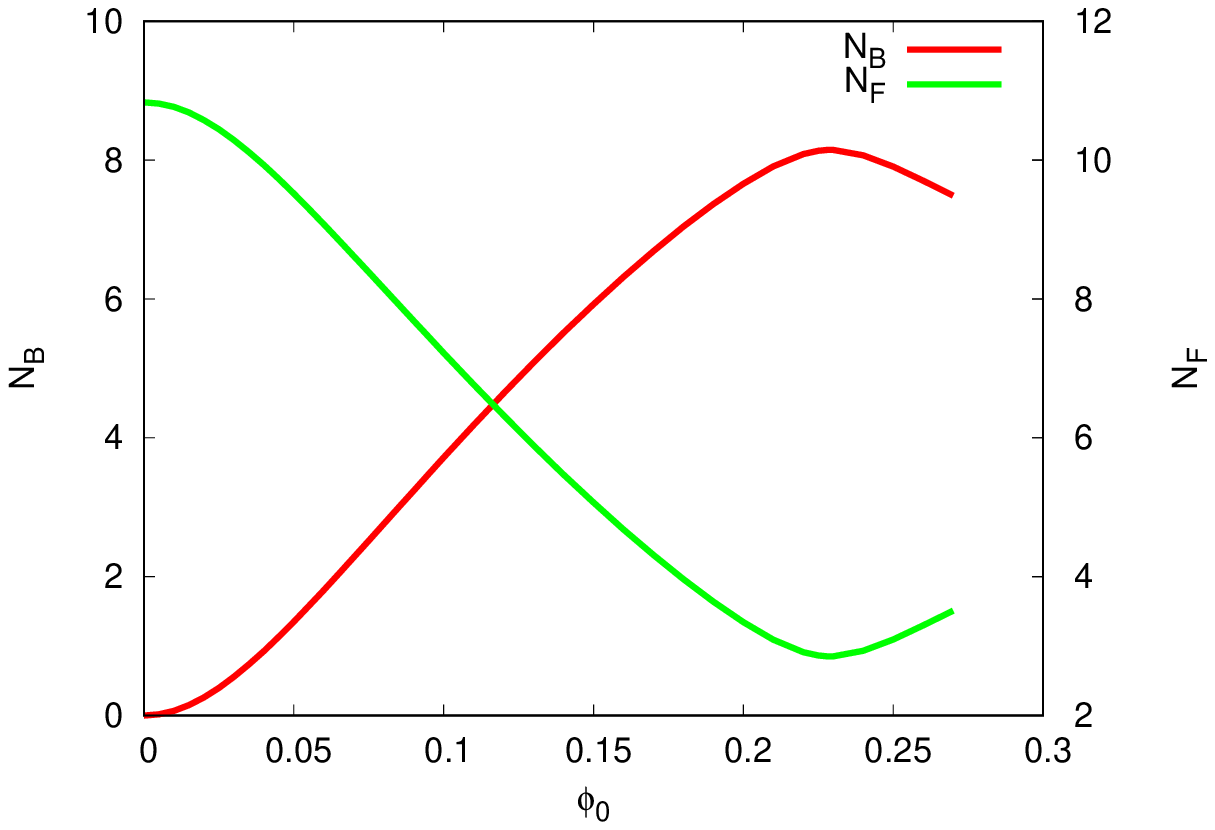}
  \includegraphics[width=0.45\textwidth]{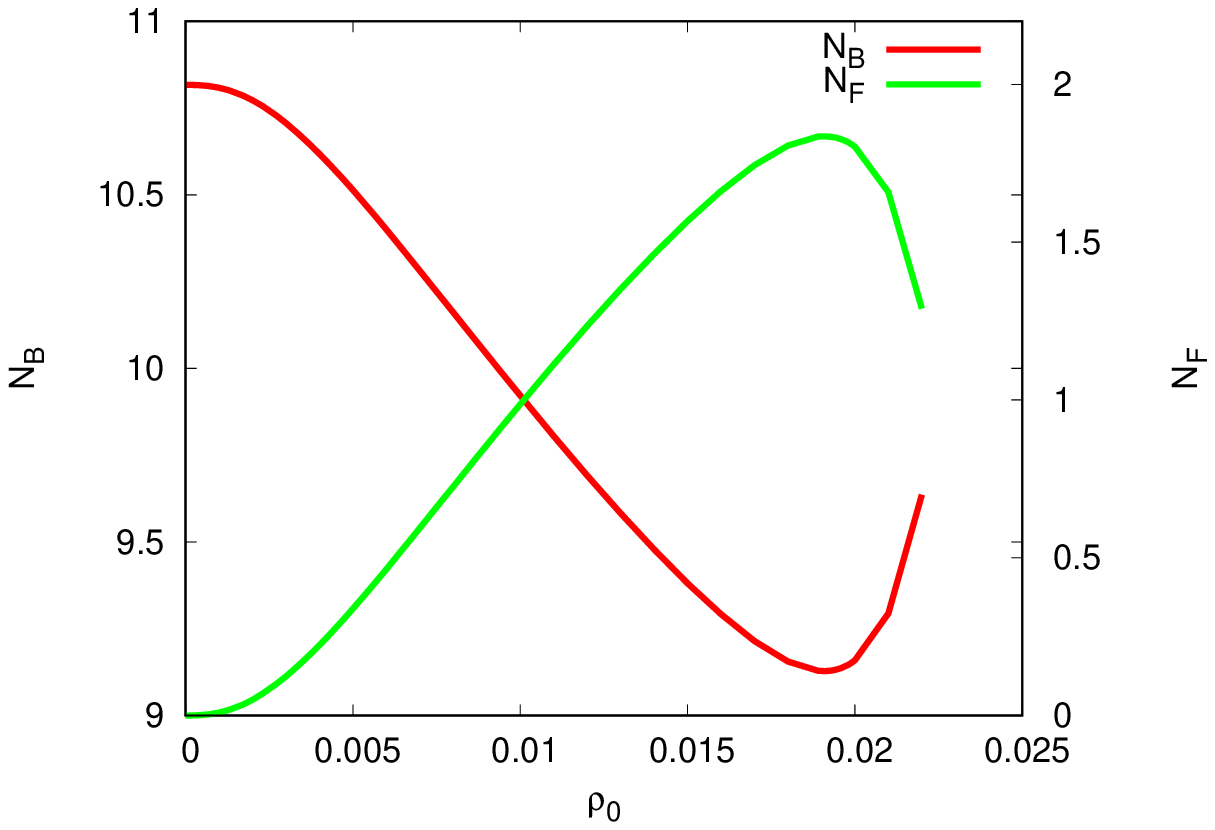}
    \caption{Number of bosons and fermions in a mixed star. (Top) It is shown how each number changes as the value of $\phi_0$ is increased, starting from a purely-fermion configuration ($\phi_0=0$). $N_F$ decreases to its minimum while $N_B$ increases to its maximum. (Bottom) Similarly, we start with a purely-boson configuration ($\rho_0=0$) and, by adding fermions, a mixed star is formed. In this case $N_B$ decreases to its minimum while $N_F$ increases to its maximum.  All plots have $\Lambda=10$ and total mass $M_T=0.83$.}
    \label{tres}
  \end{center}
\end{figure}

The same behavior is observed if we consider first a purely bosonic configuration (by providing $\phi_0$ and setting $\rho_0=0$) and subsequently adding fermions by increasing the value of $\rho_0$ from zero. In this case, $N_B$ decreases to a minimum while $N_F$ increases until reaching a maximum. This is exemplified in the two cases shown in Fig.~\ref{tres},  which were obtained with $\Lambda=10$ and fixed total mass $M_T=0.83$.

The stability analysis of mixed stars is based on this fact, and we can summarize the criterion developed in \cite{ValdezAlvarado:2012xc} as follows. The stars whose particle number is located to the left of that point where the maximum and minimum of $N_B,N_F$ coalesce are considered stable configurations, whereas
stars with particle numbers on the right of that point correspond to unstable configurations. Using this criterion, we can construct stability boundary curves on the plane $(\rho_0,\phi_0)$ by considering different values of the total mass $M_T$ (see also Fig.~\ref{totalmass}), which then split the space of possible configurations in two well defined regions.

\begin{figure}[t!]
  \begin{center}
  \includegraphics[width=0.45\textwidth]{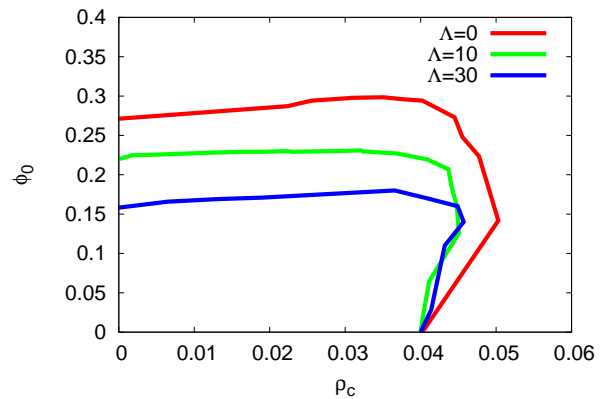}
    \caption{Boundary curves on the $(\rho_0,\phi_0)$-plane for $\Lambda=0,10,30$. The curves split the plane into two regions. Mixed stars configurations constructed from points $(\rho_0,\phi_0)$ that lie within the stable region will be stable, and unstable otherwise. As the value of $\Lambda$ increases, the stability region shrinks. See the text for details.}
    \label{Diagrama}
  \end{center}
\end{figure}

In Fig.~\ref{Diagrama} we show the boundary curves for different values of the scalar field self-interaction $\Lambda$. It must be noticed that the total mass of the configurations was varied in between two values, namely $M_\star \leq M_T \leq M_{Fc}$, where $M_\star$ is a particular value that we explain in detail in Sec.~\ref{sec:inner-struct} below. For instance, in the case $\Lambda =0$ the range of variation was $0.613 \leq M_T \leq 1.637$, and correspondingly for $\Lambda =30$ the range was $1.05 \leq M_T \leq 1.637$.

In summary, mixed stars configurations constructed from $(\rho_0,\phi_0)$ that lie within the stable region will be stable, and those outside will be unstable. It can also been noticed, in the Fig.~\ref{Diagrama}, that the stability region shrinks as $\Lambda$ increases. The boundary curves intersect the $\phi_0$-axis at different points, because the critical field value decreases for $\Lambda \neq 0$ (see Table~\ref{t1}), while they intersect the $\rho_0$-axis at the same point corresponding to $M_T=1.637$. In consequence, the maximum value of $M_T$ of stable mixed stars do not go beyond the critical mass $M_{Fc}$ of purely neutron stars.

Taking for reference the case $\Lambda =0$, there are stable and unstable configurations with total mass in the range $0.633 \leq M_T \leq 1.637$, as they correspond precisely to the curves used to determine the boundary curve shown in Fig.~\ref{Diagrama}.
Following the above reasoning, it was reported in~\cite{ValdezAlvarado:2012xc} that, for $\Lambda=0$, all mixed stars with total mass smaller than the critical
mass of purely bosonic stars, $M_T < M_c=0.633$, were stable, which was not quite precise. This is because curves corresponding to the aforementioned configurations may also have points inside and outside the stability regions.

\subsection{Inner structure of the stability region \label{sec:inner-struct}}

\begin{figure}[ht]
  \begin{center}
  \includegraphics[width=0.45\textwidth]{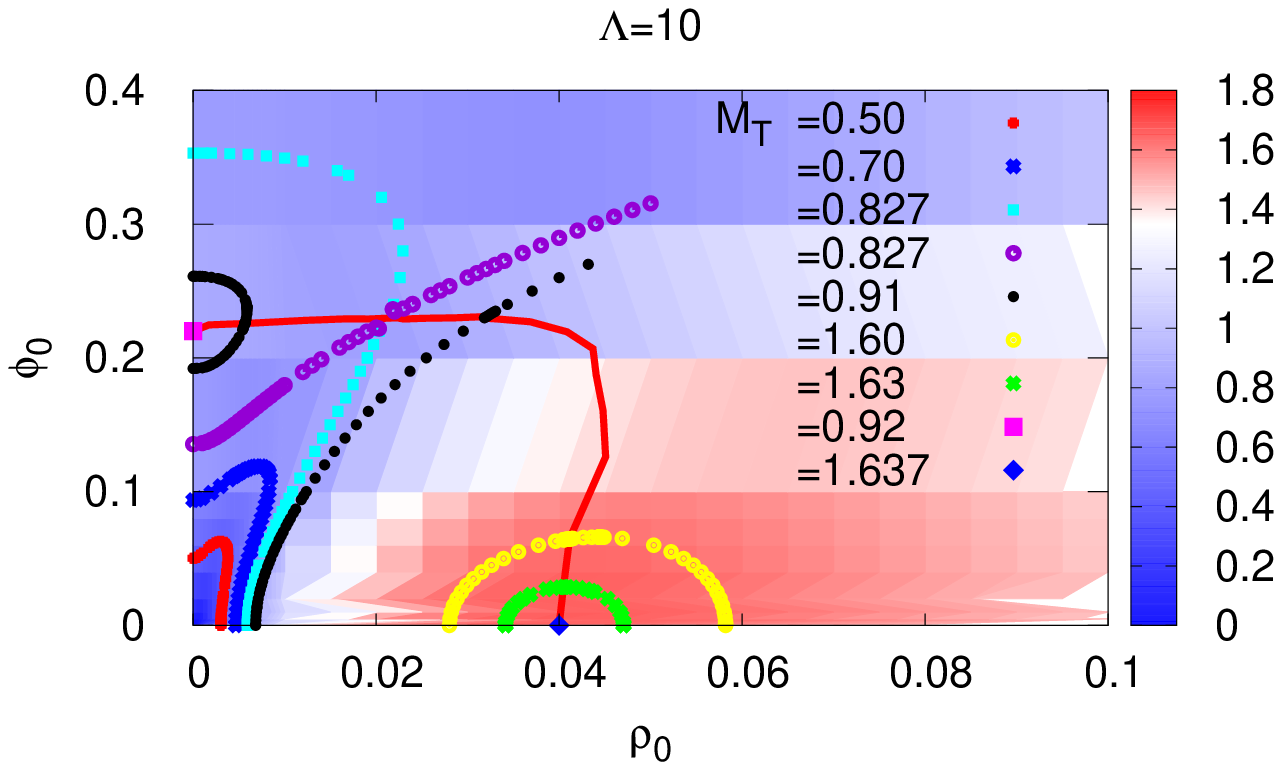}
  \includegraphics[width=0.45\textwidth]{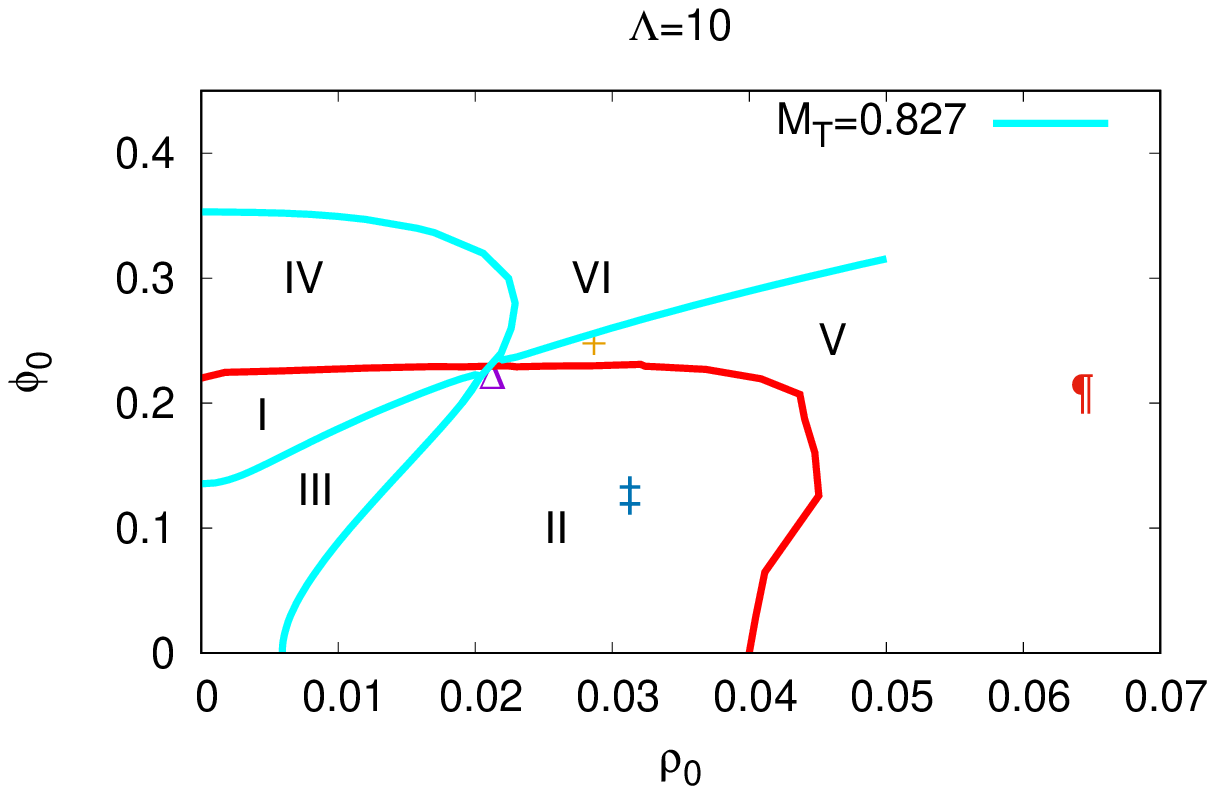}
    \caption{(Top) Stability region of mixed stars for $\Lambda=10$. Each curve corresponds to equilibrium configurations with the same total mass $M_T$. The curves reveal the stability-instability structure of the plane $(\rho_0,\phi_0)$. (Bottom) We show two curves in the $(\rho_0,\phi_0)$ plane that represent boson-fermion stars with total mass $M_T= M_\star = 0.827$ (for $\Lambda=10$). One curve starts as purely-boson star on the vertical axis (cyan curve), and the other starts as purely fermion star on the horizontal axis (purple curve). Together with the boundary red curve, they seem to play the role of separatrixes for the whole plane. In the stable region we have labeled two configurations: ($\Delta$) which represents the configuration associated with $(\rho_{0\star},\phi_{0\star})$ and $M_T=0$.$827$, ($\ddag$) corresponding to a configuration with $M_T=1$.$36$. In the unstable region we have labeled: ($+$) representing a configuration with $M_T=0$.$83$ and ($\P$) representing a configuration with $M_T=1$.$36$. The evolution of these four configurations is described in Sec.~(\ref{sec:numerical-evolution}). See the text for details.}
    \label{bifurca}
  \end{center}
\end{figure}

Similarly to the case of the boundary curves described above, one can also draw curves on the plane $(\rho_0,\phi_0)$ representing boson-fermion stars with the same total mass $M_T$, but now considering an extended range of values $0 < M_T \leq 1.637$. 
Examples of such curves are shown in the top panel of Fig.~\ref{bifurca} for the case $\Lambda=10$. For instance, all black (yellow) squares in the plot correspond to mixed stars with the same total mass $M_T=0.91$ ($M_T=0.87$), which is less than the critical boson mass $M_{Bc} =0.92$, but they are represented by two different curves: one that starts and ends at the vertical axis $\phi_0$, and another one that does the same but with respect to the horizontal axis $\rho_0$. 

Because of these behaviors, the curves must necessarily cross the boundary (red) curve, and then some of the points are within the stable region and others are outside of it. This means that there are stable and unstable mixed configurations with total mass $M_T=0.91$ ($M_T=0.87$). Other curves are shown that depict the same behavior, and the common feature in all is that they have a total mass such that $M_T > M_\star = 0.827$.
For the particular value $M_T= M_\star$, the two resultant curves (cyan and purple) meet at the same point on the boundary (red) curve, that we denote by $(\rho_{0\star},\phi_{0,\star})$, and then they represent the extreme cases of a curve that starts and ends at the same axis. Furthermore, these lines seem to determine the boundary of an inner region inside the stability one, within which all curves start on the $\phi_0$ axis and end up on the $\rho_0$ axis (or viceversa). Moreover, all configurations with total mass $M_T < M_\star$ lie inside the stability region, and then for such values of the total mass there are only stable configurations.
Thus, the stability region of boson-fermion configurations has a non-trivial inner structure, as depicted in the bottom panel of Fig.~\ref{bifurca}, with three well distinctive sub-regions that have the following properties.

\begin{itemize}
    \item Region I. The configurations have a total mass in the range $M_\star < M_T \leq M_{Bc}$, the latter value corresponding to the critical mass of purely-boson configurations. Additionally, their main characteristic is that the number of bosons is always larger than the number of fermions, $N_B > N_F$.
    \item Region II. The configurations have a total mass in the range $M_\star < M_T \leq M_{Fc}$, the latter value corresponding to the critical mass of pure-fermion configurations. Additionally, their main characteristic is that the number of fermions is always larger than the number of bosons, $N_F > N_B$.
    \item Region III. The configurations have a total mass in the range $0 < M_T < M_\star$. In contrast to the configurations in Regions I and II, this time the mixed stars can be either boson or fermion dominated, as indicated by their end points: one is on the vertical axis $\phi_0$ and another one is on the horizontal axis $\rho_0$.
    \item The boundary lines of the three inner sub-regions are: the boundary (red) curve, and the curves of the configurations with exactly the total mass $M_T=M_\star$. For these latter configurations, they can start as a purely-boson (fermion) star (or viceversa), and then change their nature by becoming a purely-fermion (boson) star. Then, they have the same number of bosons and fermions, $N_B= N_F$, when they meet at the boundary curve at the point $(\rho_{0\star},\phi_{0\star})$.
\end{itemize}

\begin{figure}[h!]
\begin{center}
\includegraphics[width=0.45\textwidth]{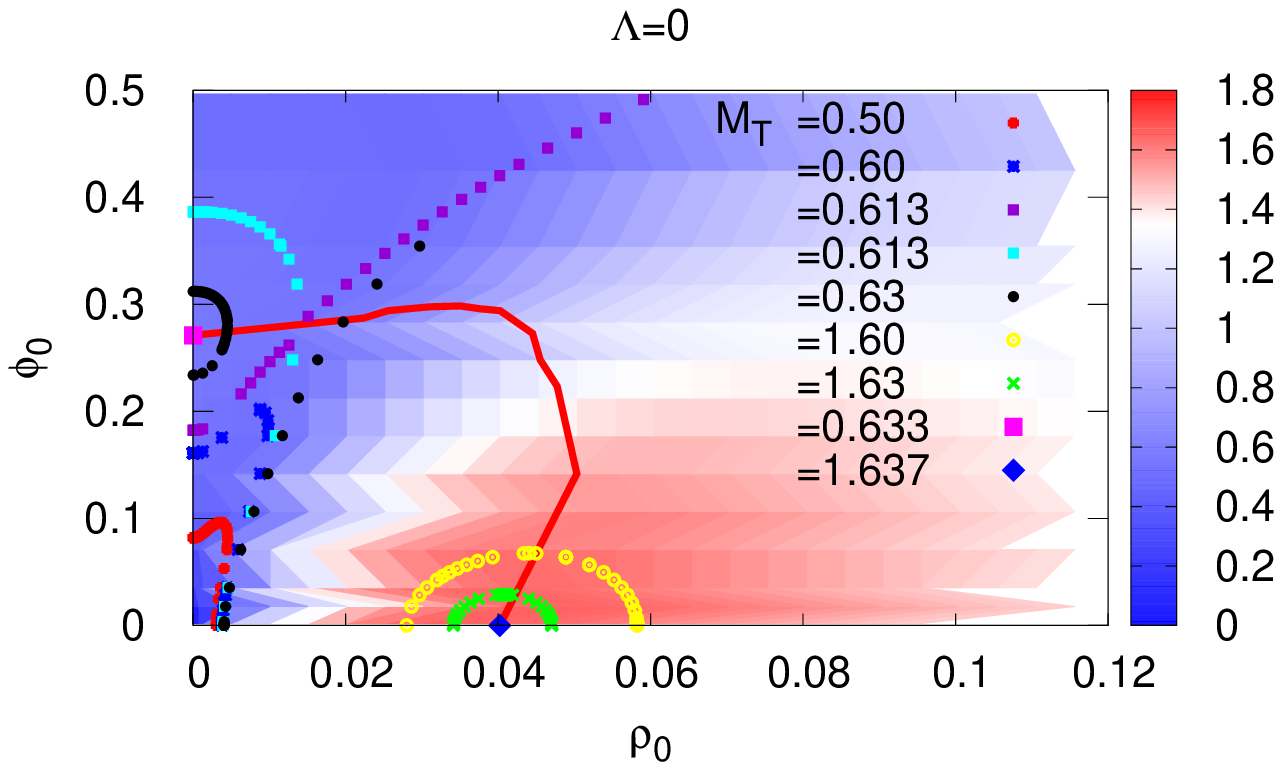}
  \includegraphics[width=0.45\textwidth]{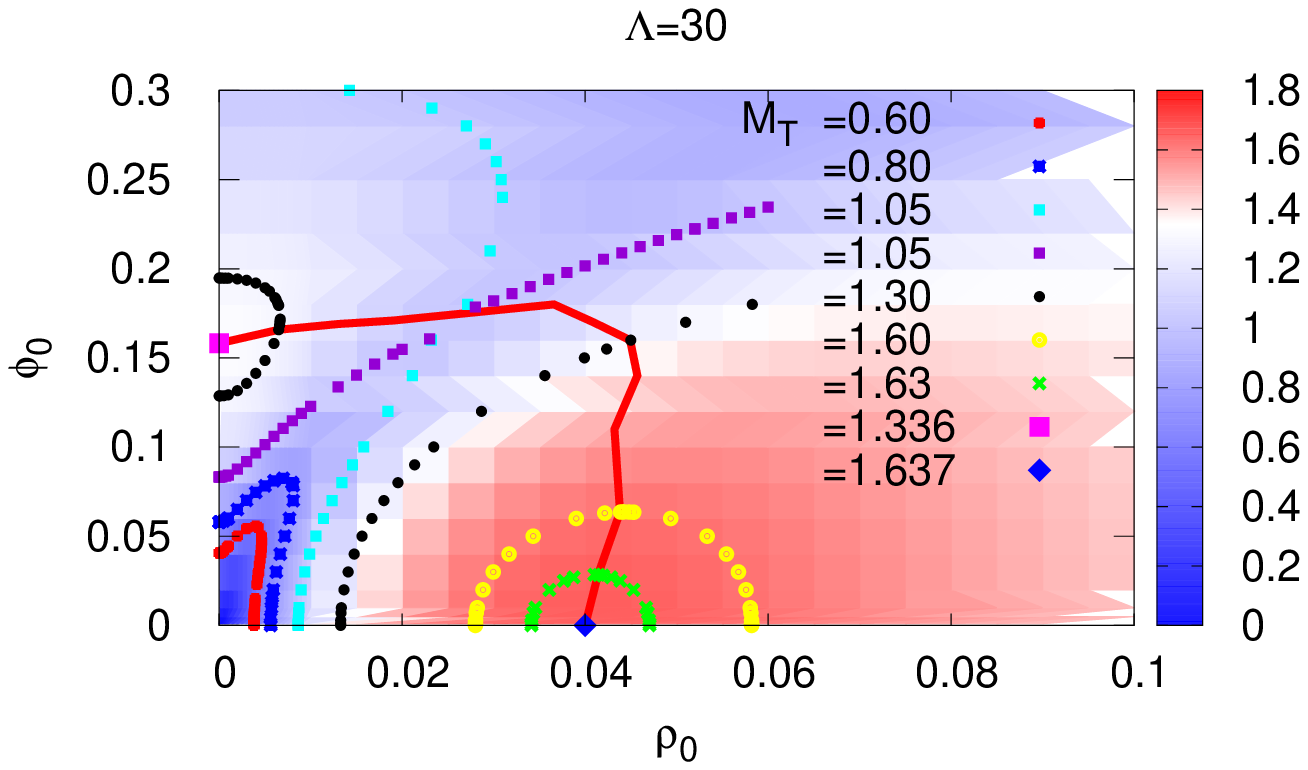}
\caption{Stability regions of mixed stars for $\Lambda=0,30$. It can be seen that they have the same structure as the case shown in Fig.~\ref{bifurca}. See the text for details.
}
\label{estabilidad}
\end{center}
\end{figure}

In our numerical experiments, we have found that the same inner structure of the stability region of mixed stars exist for any given value of the self-interaction parameter $\Lambda$, see Fig.~\ref{estabilidad}, but just with a different value of $M_\star$. For instance, for $\Lambda =0$, we find $M_\star = 0.613$, whereas for $\Lambda=30$ the corresponding value is $M_\star=1.05$. This also seems to suggest that the value of $M_\star$ increases for larger values of $\Lambda$, although one must recall that likewise the stability region becomes smaller, see Fig.~\ref{Diagrama}.

\subsection{Structure of the instability region \label{sec:inner-instruct}}

Just as in the case of the stability zone of the plane $(\rho_0,\phi_0)$, we can see in Fig.~\ref{estabilidad} that there is also a structure on the instability region, ie for the points beyond the boundary (red) curve. The sub-regions can be classified as follows.
\begin{itemize}
    \item Region IV. This region corresponds to unstable equilibrium configurations that are boson dominated, that is, $N_B > N_F$, and whose total mass is in the range $M_\star < M_T < M_{Bc}$.
    \item Region V. This region corresponds to unstable equilibrium configurations that are fermion dominated, that is, $N_B < N_F$, and whose total mass is in the range $M_\star < M_T < M_{Fc}$.
    \item Region VI. This region corresponds to unstable equilibrium configurations that can be either boson or fermion dominated. The total mass of these configurations varies in a narrow range, which depends on the value of the self-interaction $\Lambda$, but in general around and above the critical value $M_\star$.
\end{itemize}
Although we have not been able to do a thorough exploration of all possible solutions, we can see that the structure of the unstable region we have just described is consistent with that of the stable region explained in Sec.~\ref{sec:inner-struct} above. In that consistency the special value $M_\star$ plays a central role and seems to be the common element of the overall stability and instability regions.

\subsection{Numerical Evolution
\label{sec:numerical-evolution}}

To corroborate the stability analysis described in Sec.~\ref{sec:equilibrium-conf} above, we will perform the numerical evolution of four equilibrium configurations with $\Lambda=10$: the configuration associated to $(\rho_{0\star},\phi_{0\star})$, that is the configuration corresponding to the intersection between the curves with $M_{\star}$ and the boundary stability curve; one configuration in region $II$ with $M_T=1$.$63$; and two configurations in region $V$ with $M_T=0.83$ and $M_T=1$.$63$. 

These four configurations have been labeled in the bottom panel of Fig.~\ref{bifurca}, respectively, by $\Delta$, $\ddag$, $+$ and $\P$. The numerical results were obtained from the evolution of the Einstein-Klein-Gordon-Hydrodynamics equations
~\eqref{eq:ekghd}, and~\eqref{eq:auxiliar}-\eqref{eq:forz3}, using the same numerical methods described in~\cite{ValdezAlvarado:2012xc}. 

\begin{figure}[ht]
  \begin{center}
  \includegraphics[width=0.45\textwidth]{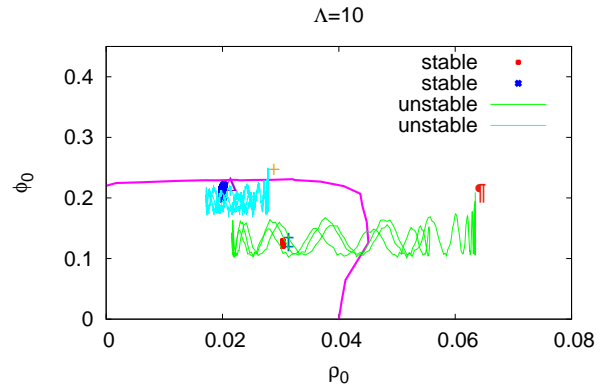}
    \caption{Behavior of the central values of fluid density and peaks of the oscillatory scalar field, ($\rho_0,\phi^{max}_0$), for the four configurations labeled in the caption of fig.~(\ref{bifurca}). The configuration represented by ($\Delta$) and ($\ddag$) are stable and remain in the same state, whereas configurations represented by ($\P$) and ($+$) are unstable and migrate toward the stable regions.}
    \label{estabilidadevol}
  \end{center}
\end{figure}

Figure~\ref{estabilidadevol} shows the behavior of the central values of the fluid density and scalar field of the four mixed stars. As we can see, the configurations in the intersection and in region $II$ are stables because remain in the same state during evolution. While the configurations in region $V$ are, indeed, unstable because migrate toward the stable regions.   

\begin{figure}[ht]
\begin{center}
\includegraphics[width=0.45\textwidth]{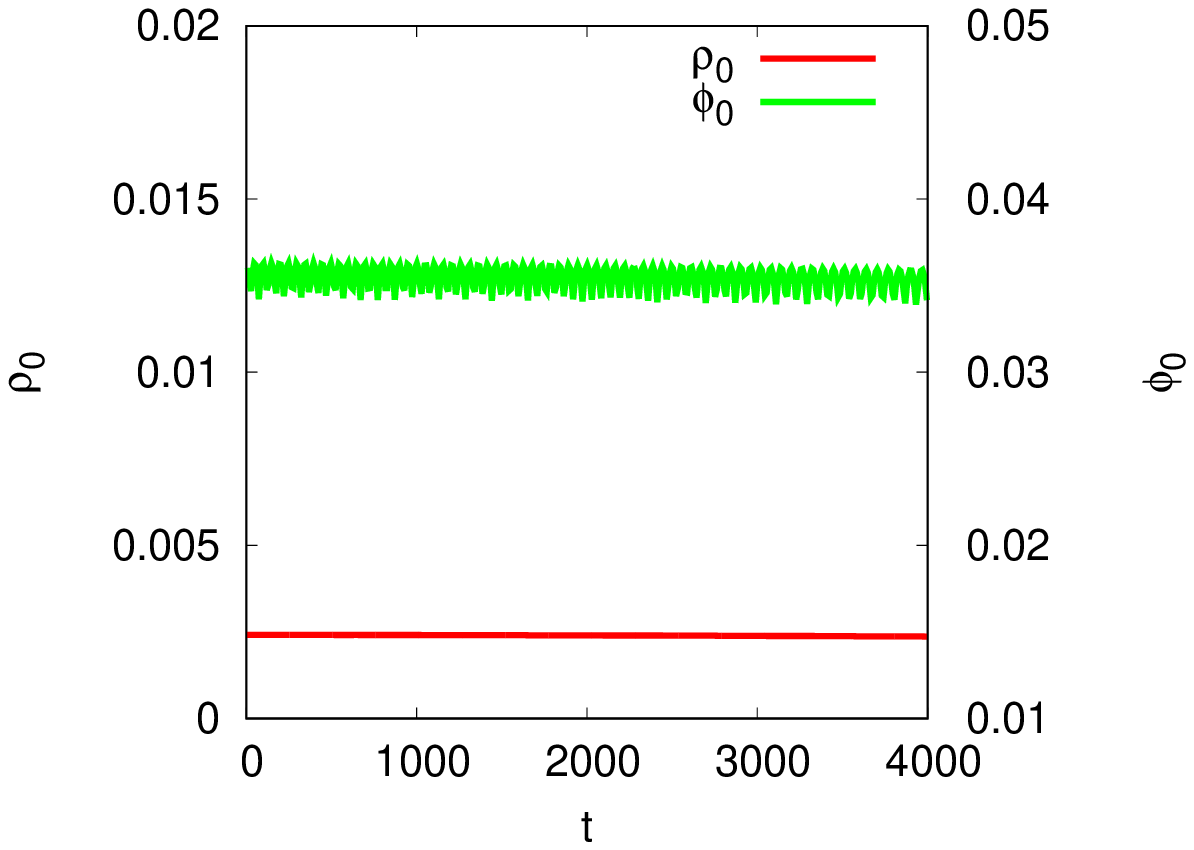}
  \includegraphics[width=0.45\textwidth]{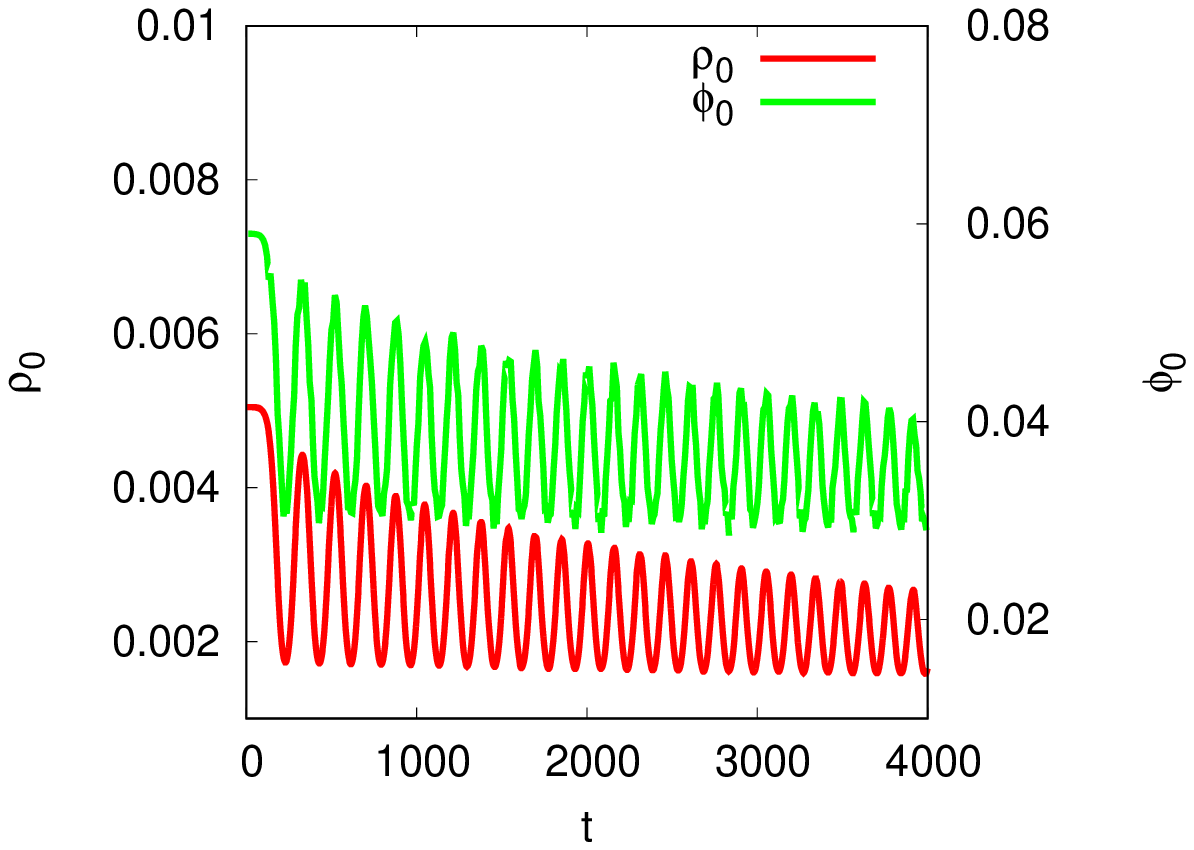}
\caption{Evolution of mixed stars. (Top) Stable configuration. The central values of the density and the (peaks of the oscillatory) scalar field remain very close to their initial values, suggesting that the star is stable against perturbations. (Bottom) Unstable configuration. The central values of the density and (the peaks of) the scalar field depart quickly from their initial values, indicating that the star is unstable. The evolution becomes nonlinear and describes the migration of the star from the unstable to the stable branch. 
}
\label{rhophievol}
\end{center}
\end{figure}

In Fig.~\ref{rhophievol} we show the evolution, as a function of time, of the central value of the perfect fluid and scalar field in the case of the stable and unstable configuration with total mass $M_T=1$.$63$. For the stable configuration (labeled with ($\ddag$)), the scalar field oscillates with its characteristic eigenfrequency, while the fluid density oscillates slightly around its initial state due to the perturbation introduced by the numerical truncation errors. These quantities remain very close to their initial value, indicating that the configuration is stable. On the other hand, the unstable configuration (labeled with ($\P$)), presents remarkable variations in the amplitudes of the oscillation of scalar field and the fluid density, then, the star is eventually migrating from the unstable to the stable branch.

\begin{figure}[ht]
  \begin{center}
  \includegraphics[width=0.45\textwidth]{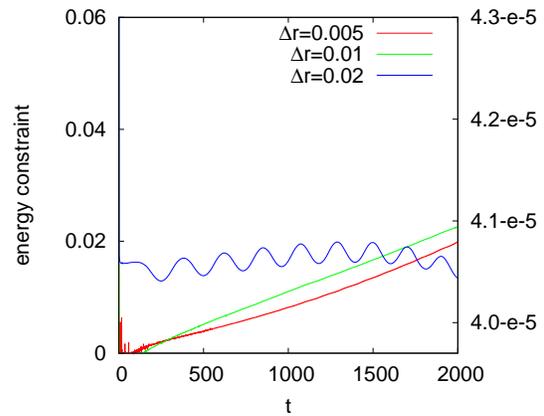}
    \caption{The energy constraint for three different resolutions: $\Delta x=0.005$ (red), $\Delta x=0.01$ (green), and $\Delta x=0.02$ (blue). The scale of red and green lines is indicated on the right side of the graph. We can see a second order convergence.}
    \label{hamilton}
  \end{center}
\end{figure}

 To revise the accuracy of our numerical calculations, we monitored the energy constraint~(\ref{eq:hamilton-constraint}) evolving a configuration with three different spatial resolution, $\Delta r=(0$.$005, 0$.$01,0$.$02)$ at $r_{max}=600$ and $t_{max}\approx 2000$. This configuration correspond to the $\rho_0=0.005$, $\phi_0=0.05$ and $\Lambda=10$. In Fig.~\ref{hamilton} we can see that the energy constraint remain small during the evolution and converges to zero.
 
 \textbf{}
%
\section{Final remarks and discussion \label{sec:final-rem}}
In this paper we have applied the criterion
developed in~\cite{ValdezAlvarado:2012xc} to determine the stability of fermion-boson stars with quartic self-interaction for the bosonic part. Using this criterion, stability boundary curves in the $(\rho_0,\phi_0)$ plane were constructed. We have again found a main curve that splits the plane in two well defined regions, one that contains stable configurations and another with unstable ones. It turns out that as the value of the self-interaction parameter $\Lambda$ increases, the stability region shrinks. It was also shown that the maximum value of the total mass of mixed stars do not go beyond the critical mass of purely neutron stars, a results that stands for non-zero values of $\Lambda$.

Additionally, we have been able to unravel the structure of both the stable and unstable regions, by means of the special curves corresponding to equilibrium configurations with total mass $M_\star$. These curves are special, they meet at the boundary curve at the same point, which means that at such point the equilibrium configuration has the same number of bosons and fermions. 

We have then argued that the value of $M_\star$ is a true discriminant for the existence of stable equilibrium configurations. Our study suggests that equilibrium configurations with total mass $M_T < M_\star$ are intrinsically stable (whether boson or fermion dominated), whereas one can construct either stable or unstable configurations in the opposite case $M_T > M_\star$ (again, whether boson or fermion dominated). The value of $M_\star$ depends on the value of the self-interaction parameter $\Lambda$, but its role as discriminant for intrinsically stable configurations is always the same.

To assess the stability criterion, we performed the numerical evolution of the fully nonlinear equations of motion for some typical configurations. For the stable configuration, the central values of the scalar field and the fluid density remain constant in time during the numerical evolution, while the unstable star migrates to a stable configuration. Our results on the numerical evolution of the boson-fermion stars coincide with the recent study in~\cite{giovanni2020dynamical}, which is focused on the formation of this type of stars and also considers a self-interaction term in the bosonic sector. It is shown there that the resultant equilibrium configurations are stable under more general conditions, which further validates our stability analysis in Sec.~\ref{sec:stability-analysis} above.

Our study considers the effects coming from the addition of a bosonic self-interaction, but it would be interesting to study the effects of varying the polytropic and adiabatic index on the stability of mixed stars. This is a topic of an ongoing research that we expect to report elsewhere.

%
\begin{acknowledgements}
We want to thank Dana Alic and Carlos Palenzuela for allowing us to use their numerical code for the evolution of the fermion-boson equilibrium configurations considered in the main text. We thank Cinvestav Physics Departament for letting us employ its computer facilities and acknowledge Jazhiel Chacón Lavanderos for technical support. SV-A acknowledges partial support from Consejo Nacional de Ciencia y Tecnolog\'ia (CONACyT), under a research assistant grant and Programa para el Desarrollo Profesional Docente (PRODEP), under 4025/2016RED project. RB acknowledges partial support from CIC-UMSNH. LAU-L was partially supported by Programa para el Desarrollo Profesional Docente; Direcci\'on de Apoyo a la Investigaci\'on y al Posgrado, Universidad de Guanajuato, research Grant No. 036/2020; CONACyT M\'exico under Grants No. A1-S-17899, 286897, 297771; and the Instituto Avanzado de Cosmolog\'ia Collaboration.
\end{acknowledgements}

%
%

\appendix

\section{The evolution equation of space-time}
\label{app:Einsten-equations}

The $Z3$ formulation of the Einstein equation, in spherical
symmetry~\cite{Bernal:2009zy}, introduces the following auxiliary variables in order to obtain a first order system of equations 
\begin{eqnarray}
A_r &=& \frac{\partial_r\alpha}{\alpha}\, , \,\,\,
D_{rr}{}^{r}=\frac{g^{rr}}{2}\partial_rg_{rr}\, ,\,\,\,
D_{r\theta}{}^{\theta}=\frac{g^{\theta\theta}}{2}\partial_rg_{\theta\theta}\,,
\nonumber \\
K_{r}{}^{r} &=& -\frac{1}{2 \alpha} \frac{\partial_{t}g_{rr}}{g_{rr}} \, , \,\,\,
K_{\theta}{}^{\theta} = -\frac{1}{2 \alpha}
     \frac{\partial_{t}g_{\theta\theta}}{g_{\theta\theta}}\, ,
     \label{eq:auxiliar}
\end{eqnarray}

For the lapse function, we use the harmonic slicing condition 
\begin{equation}
    \partial_t\alpha = -\alpha^2 trK
    \label{eq:slicing}
\end{equation}
where $trK=K^{r}{}_{r}+2K^{\theta}{}_{\theta}$ is the trace of the extrinsic curvature. The system is regularized at the origin using the following transformation of the momentum constraint: 
\begin{eqnarray}
\tilde{Z_{r}} &=& Z_{r} +
\frac{1}{4r}\left(1-\frac{g_{rr}}{g_{\theta\theta}}\right) \, ,
\end{eqnarray}
this transformations avoids problems at $r \rightarrow 0$. Then, in terms of theses variables, the evolution equation of the geometry are written as
\begin{widetext}
\begin{subequations}
\label{eq:forz3}
\begin{eqnarray}
  \partial_{t}A_{r} &=& -\partial_{r}[\alpha \, trK] \,, \label{Ar} \\
  \partial_{t}D_{rr}{}^{r} &=& -\partial_{r}[\alpha \, K_{r}{}^{r}] \,, \label{drr} \\
  \partial_{t}D_{r\theta}{}^{\theta} &=& -\partial_{r}[\alpha \,
  K_{\theta}{}^{\theta}] \,, \label{dthetatheta} \\
  \partial_{t}Z_{r} &=& -\partial_{r}[2 \, \alpha \, K_{\theta}{}^{\theta}] +
  2\alpha\Big\{(K_{r}{}^{r}-K_{\theta}{}^{\theta})\Big(D_{r\theta}{}^{\theta}
+ \frac{1}{r} \Big) - K_{r}{}^{r} \Big[ Z_{r} + \frac{1}{4r} \Big( 1 -
\frac{g_{rr}}{g_{\theta\theta}} \Big) \Big]  \nonumber\\
 & & + A_{r}K_{\theta}{}^{\theta}+\frac{1}{4r}\frac{g_{rr}}{g_{\theta\theta}}
  (K_{\theta}{}^{\theta}-K_{r}{}^{r}) -4\pi S_{r}\Big\} \, , \label{zr} \\
\partial_{t}K_{r}{}^{r} &=& -\partial_{r}\Big[\alpha g^{rr}\Big(A_r+
\frac{2}{3}D_{r\theta}{}^{\theta}-\frac{4}{3}Z_{r}\Big)\Big]+
\alpha\Big\{(K_{r}{}^{r})^{2}+\frac{2}{3}K_{\theta}{}^{\theta}(K_{r}{}^{r}-K_{\theta}{}^{\theta}) \nonumber\\
 & & - g^{rr}D_{rr}{}^{r}A_{r} +\frac{1}{3r}[g^{rr}(D_{rr}{}^{r}-A_{r}-4Z_{r})+g^{\theta\theta}
 (D_{r\theta}{}^{\theta}-A_{r})] \nonumber\\
 & & + \frac{2}{3}g^{rr}\Big[Z_{r}+\frac{1}{4r}\Big(1-\frac{g_{rr}}
 {g_{\theta\theta}}\Big)\Big](2D_{rr}{}^{r}-2D_{r\theta}{}^{\theta}-A_{r}) \nonumber\\ 
 & & - \frac{2}{3}g^{rr}\Big(D_{r\theta}{}^{\theta}+
\frac{1}{r}\Big)(D_{rr}{}^{r}-A_{r})+
8\pi\Big(\frac{\tau}{6}-\frac{S_{r}{}^{r}}{2}+S_{\theta}{}^{\theta}\Big)\Big\}
\, ,\label{krr}\\
\partial_{t}K_{\theta}{}^{\theta} &=& -\partial_{r}\Big[\alpha
g^{rr}\Big(-\frac{1}{3}D_{r\theta}{}^{\theta}+\frac{2}{3}Z_{r}\Big)\Big]+
\alpha\Big\{\frac{1}{3}K_{\theta}{}^{\theta}(-K_{r}{}^{r}+4K_{\theta}{}^{\theta})\nonumber\\
 & & +  \frac{1}{6r}[g^{rr}(A_{r}-2D_{rr}{}^{r}-4Z_{r})+g^{\theta\theta}(A_{r}-2D_{r\theta}{}^{\theta})] \nonumber\\
 & & - \frac{2}{3}g^{rr}\Big[Z_{r}+\frac{1}{4r}\Big(1-
\frac{g_{rr}}{g_{\theta\theta}}\Big)\Big](D_{rr}{}^{r}-D_{r\theta}{}^{\theta}-2A_{r})\nonumber\\
 & & + \frac{1}{3}g^{rr} \Big(D_{r\theta}{}^{\theta}+\frac{1}{r} \Big)(D_{rr}{}^{r}-4A_{r})+ 
8\pi\Big(\frac{\tau}{6}-\frac{S_{r}{}^{r}}{2}+S_{\theta}{}^{\theta}\Big)\Big\}
\, ,\label{kthetatheta}
\end{eqnarray}
\end{subequations}
\end{widetext}
where $Z_{r}$ is the vector associated with the Z3 formulation, and total matter terms are given by
\begin{subequations}
\begin{eqnarray}
  \tau &=& \frac{1}{2}(g^{rr}\phi^{*}_{t}\phi_{t}+g^{rr}\phi^{*}_{r}
  \phi_{r}+V(\phi))+U\, , \\
  S_{r} &=&-\frac{1}{2}[\sqrt{g^{rr}}\phi^{*}_{t}\phi_{r}+\sqrt{g^{rr}}
  \phi_{t}\phi^{*}_{r}] + \tilde{S}_{r}\, , \\
  S_{r}{}^{r} &=&\frac{1}{2}[g^{rr}\phi^{*}_{t}\phi_{t}+g^{rr}
  \phi^{*}_{r}\phi_{r}-V(\phi)] + \tilde{S}_{r}{}^{r} \,
  , \\
  S_{\theta}{}^{\theta} &=& \frac{1}{2}[g^{rr}\phi^{*}_{t}\phi_{t}-g^{rr}
  \phi^{*}_{r}\phi_{r}-V(\phi)]+\tilde{S}_{\theta}{}^{\theta}\,.
\end{eqnarray}
\end{subequations}

In order to test the accuracy of the numerical calculations, we use the Hamiltonian constraint which is defined as
\begin{eqnarray}
H &=& \frac{2}{g_{rr}} \Big\{ -2 \partial_i D_{r \theta}{}^{\theta} - 3 D_{r \theta}{}^{\theta}
\Big(D_{r \theta}{}^{\theta} + \frac{2}{r} \Big) \nonumber\\
& & + g_{rr} K_{\theta}{}^{\theta} (K_{\theta}{}^{\theta} + 2 K_{r}{}^{r}) 
- \frac{(1 - g_{rr} g^{\theta\theta})}{r^2} \nonumber\\
& & + 2 D_{rr}{}^{r} \Big(\frac{1}{r} 
+ D_{r\theta}{}^{\theta} \Big) - 8 \pi g_{rr} \tau \Big\}. \label{eq:hamilton-constraint}
\end{eqnarray}
\section{The transformation from conserved to primitive quantities}
\label{app:conserved-primitive}

The conserved quantities are defined as
\begin{equation}\label{eq:conserved}
   D = \rho_{0} W\,, \quad  U = h W^{2} - P \,, \quad 
   \tilde{S}_{r} = h W^2 v_{r} \, ,
\end{equation}
where $h=\rho(1+\epsilon)+ P$ is the enthalpy and $W=1/\sqrt{1-v^rv_r}$ is the Lorentz factor. The spatial projections of the stress-energy tensor are given by
\begin{eqnarray}\label{eq:rhs_conserved}
   \tilde{S}_{r}{}^{r} &=& h W^2 v_{r} v^{r} + P \,, \quad 
   \tilde{S}_{\theta}{}^{\theta} = P \,.
   \label{eq:projections}
\end{eqnarray}

Then, the primitives variables. $\{\rho\, , P\, , v_r\,, \epsilon\}$ must be calculated after each time integration of the equations of motion, because they are necessary to calculated the projections of the strees-energy density~(\ref{eq:projections}). This is not trivial, mainly because the enthalpy $h$, and the
$W$, are defined as functions of the
primitives. 

We are adopting a recovery procedure which consists in the following steps:
\begin{enumerate}
\item From the first thermodynamics law for adiabatic processes, it
  follows that
\begin{equation}\label{presion}
P = (\Gamma -1) \rho\epsilon \, .
\end{equation}
Substituting the definition of the entalphy in the equation of state above, 
we write the pressure as a function of the conserved quantities and
the unknown variable $x = h W^2$. 

\item Using the previous step, the definition of $U$ becomes:
\begin{eqnarray}\label{eq:forx}
   U &=& hW^2 - P\nonumber\\ 
   &=& hW^2 -\frac{(\Gamma -1)}{\Gamma}(h-\rho)\nonumber\\
   &=& hW^2\Big(1-\frac{\Gamma-1}{\Gamma}\Big)+
   \frac{\Gamma-1}{\Gamma}\rho \, ,
\end{eqnarray}
where $\Gamma$ is the adiabatic index corresponding to an ideal gas.

\item Then, the function
\begin{eqnarray}
f(x) &=& \left(1 - \frac{\Gamma - 1}{W^2 \Gamma} \right) x +
\frac{D (\Gamma - 1)}{W \Gamma} - U,
\end{eqnarray}
must vanish for the physical solutions. The roots of the function
$f(x)=0$ can be found numerically by means of an iterative
Newton-Raphson solver, so that the solution at the $n+1$-iteration can
be computed as
\begin{equation} 
x_{n+1} = x_n - \frac{f(x_n)}{f'(x_n)},
\end{equation} 
where $f'(x_n)$ is the derivative of the function $f(x_n)$. The
initial guess for the unknown $x$ is given in the previous time step.

\item After each step of the Newton-Raphson solver, we update the
  values of the fluid primitives as
\begin{equation}\label{eq:primitive}
   \rho = D/ W\,, \quad P = x - U \,, \quad 
   v_{r} = \tilde{S}_{r}/x \, , \\ 
\end{equation}
where $W^2=x^2/(x^2-\tilde{S}^r\tilde{S}_r)$.

\item Iterate steps 3 and 4 until the difference between two
  successive values of $x$ falls below a given threshold value of the
  order of $10^{-10}$.
\end{enumerate}
\vspace{0.5cm}
\section{Limit case: $\Lambda\rightarrow\infty$}
\label{app1:limite}
To study the limit $\Lambda \gg 1$ we proceed as in Ref.~\cite{Colpi:1986ye}. We start by defining the following non-dimensional variables
\begin{equation}
x_* = x \, \Lambda^{-1/2}\, , \quad  \phi_* = \phi \, \Lambda^{1/2}\, , \quad  M_\ast =M  \, \Lambda^{1/2} \, . \label{var:limit}
\end{equation}
Substituting the variables~\eqref{var:limit} in the system of equations~\eqref{eq:ekghd} and ignoring terms $O(\Lambda^{-1})$, we obtain an algebraic equation for the scalar field, $\phi_\ast = (S-1)^{1/2}$, where $S \equiv \Omega^2/\alpha^2$. Additionally, if we write $a^2 = \left( 1- 2M/x \right)^{-1} = \left( 1- 2M_\ast/x_\ast \right)^{-1}$, the equations of motion turn out to be
\begin{subequations}
\label{eq:limit}
\begin{eqnarray}
M^\prime_\ast &=& \frac{x^2_\ast}{4} \Big[ (3S+1)(S-1)+ 4 \Lambda \, \rho (1+\epsilon) \Big] \, , \\
S^\prime &=& -S a^2 \left(\frac{2M_\ast}{x^2_*}+\frac{x_*}{2}\Big[(S-1)^2 + 4 \Lambda \, p \Big]\right) \, , \\
p^\prime &=& \frac{S^\prime}{2S} [\rho (1+\epsilon) + p] \,,
\end{eqnarray}
\end{subequations}
where now a prime denotes derivative with respect to $x_\ast$.

\begin{figure*}[ht]
\begin{center}
\includegraphics[width=0.45\textwidth]{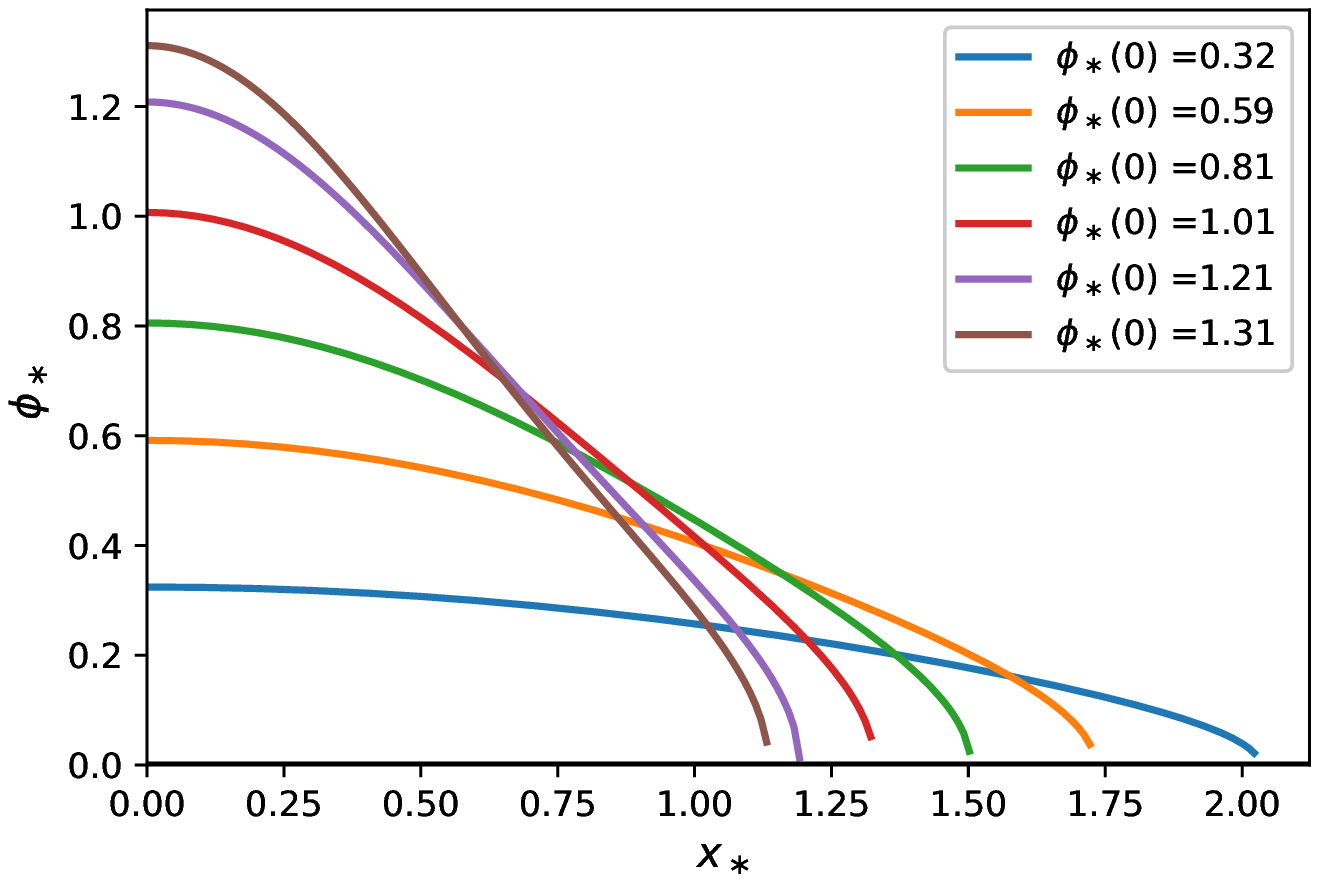}
\includegraphics[width=0.45\textwidth]{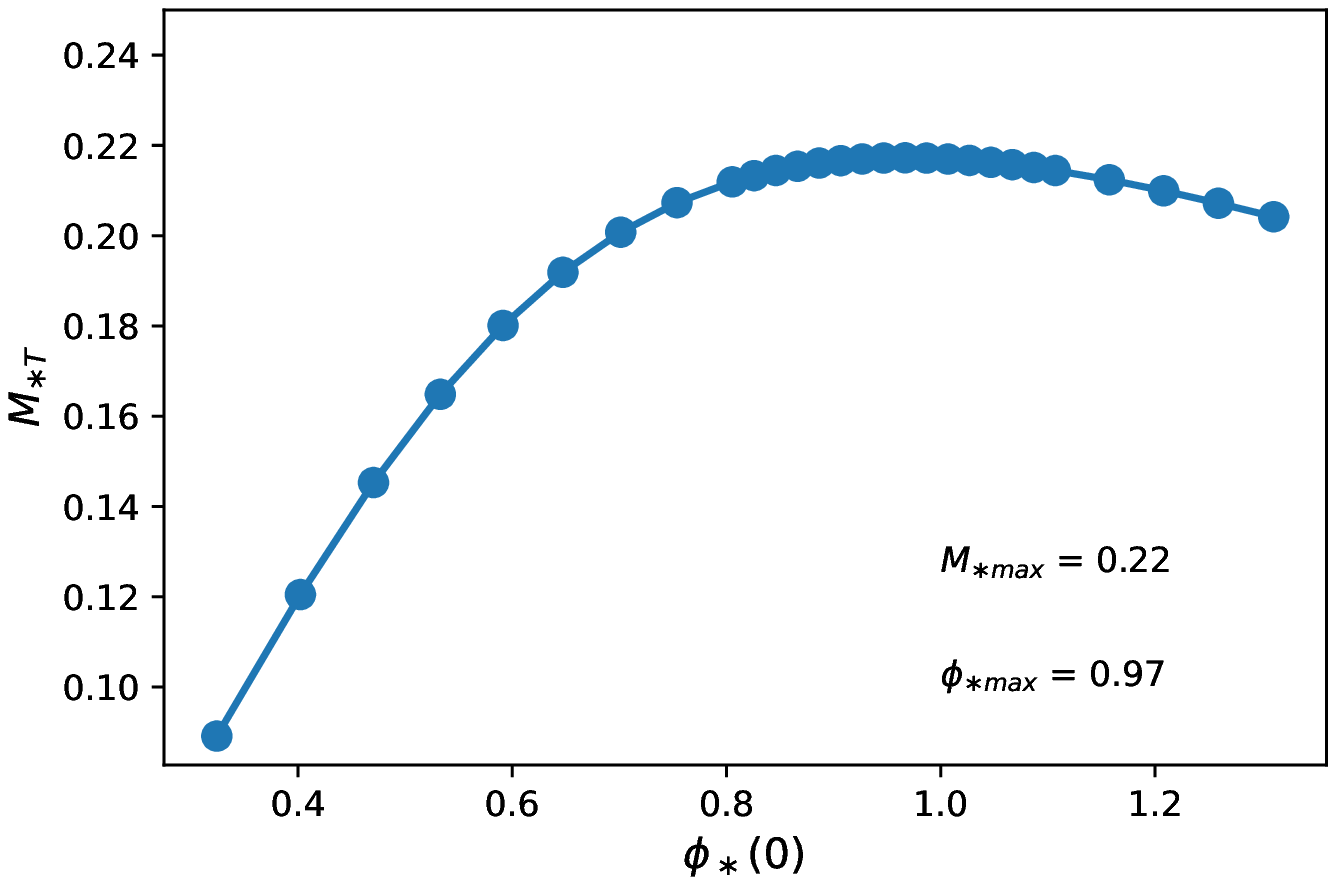}
\caption{\label{fig:colpi} Single boson star in the limit $\Lambda \gg 1$. (Left) The field profile $\phi_\ast$ as obtained from Eqs.~\eqref{eq:limit}, classified in terms of the central values $\phi_\ast(0)$. (Right) The total mass $M_{\ast T}$ as a function of the central value $\phi_\ast (0)$. The maximum total mass is $M_{\ast max} = 0.22$ for $\phi_{\ast max} = 0.97$. See text for more details.}
\end{center}
\end{figure*}

In the standard case of a single boson star one can see that $\Lambda$ does not appear in the equations of motion, and then one obtains general solutions that are valid for any value of $\Lambda$. We show in the left panel of Fig.~\ref{fig:colpi} the numerical solution of $\phi_\ast$ as a function of $x_\ast$ as obtained from Eq.~\eqref{eq:limit}. The solutions are labeled in terms of the central value $\phi_\ast(0) = S(0) = \Omega^2/\alpha^2(0)$; notice that in the limit $\Lambda \gg 1$ it is just enough to set the central value $\phi_\ast(0)$ to determine the full solution.

In the right panel of Fig.~\ref{fig:colpi} we show the total mass $M_{\ast T}$ as a function of the central value $\phi_{\ast}(0)$. Notice that the maximum total mass for the stability of the boson star, in the limit $\Lambda \gg 1$, is $M_{\ast max} =0.22$, which is in agreement with the results in~\cite{Colpi:1986ye}. Thus, the maximum mass for stability is $M_{max} \simeq 0.22 \, \Lambda^{1/2}$ whereas $\phi_{max} \simeq 0.97 \, \Lambda^{-1/2}$.

In contrast, for a boson-fermion star $\Lambda$ is explicitly present and enhances the contribution of the perfect fluid in the scaled equations of the metric quantities $M_\ast$ and $S$. Although one can make a similar transformation of the fluid quantities (eg $\rho_\ast = \Lambda \, \rho$), this will distort the properties of the fermionic fluid. Hence, the overall lesson is that Eqs.~\eqref{eq:limit} provide boson-fermion solutions in the limit $\Lambda \gg 1$ for any pair of values $(\rho(0),\phi_\ast(0))$, similarly to the equilibrium configurations of the full system~\eqref{eq:ekghd}.

Another consequence of the limit $\Lambda \gg 1$ is that the critical mass of a single boson star can be larger than that of a perfect fluid; in our case, this happens if $\Lambda > (1.637/0.22)^2 \simeq 55$. A direct implication is that one can have stable boson-fermion star with a mass larger than the critical mass of single fermion star. As for the overall stability, the only change in the analysis in Sec.~\ref{sec:stability-analysis} would be the interchange $M_{Fc}$ by $M_{Bc}$, and vice versa, if it is the case that $M_{Bc} > M_{Fc}$.

%
\bibliography{paper}

\end{document}